\documentclass[aps,10pt,notitlepage,nofootinbib,showpacs,superscriptaddress,longbibliography]{revtex4-1}

\expandafter\let\csname equation*\endcsname\relax
\expandafter\let\csname endequation*\endcsname\relax
\expandafter\let\csname eqnarray*\endcsname\relax
\expandafter\let\csname endeqnarray*\endcsname\relax
\usepackage{amsmath}
\usepackage{amssymb}
\usepackage{graphicx}
\usepackage{xcolor}

\usepackage[hidelinks]{hyperref}

\renewcommand{\rm}{\mathrm}

\begin{document}
\title{Dynamical response of Bose-Einstein condensates to oscillating gravitational fields}
\author{Dennis R\"atzel}
\email{dennis.raetzel@univie.ac.at}
\affiliation{Faculty of Physics, University of Vienna, Boltzmanngasse 5, 1090 Vienna, Austria}

\author{Richard Howl}
\affiliation{Faculty of Physics, University of Vienna, Boltzmanngasse 5, 1090 Vienna, Austria}

\author{Joel Lindkvist}
\affiliation{Faculty of Physics, University of Vienna, Boltzmanngasse 5, 1090 Vienna, Austria}

\author{Ivette Fuentes}
\affiliation{School of Mathematical Sciences, University of Nottingham, University Park, Nottingham NG7 2RD, UK}
\affiliation{Faculty of Physics, University of Vienna, Boltzmanngasse 5, 1090 Vienna, Austria}
\begin{abstract}

A description of the dynamical response of uniformly trapped Bose-Einstein condensates (BECs) to oscillating external gravitational fields is developed, with the inclusion of damping. Two different effects that can lead to the creation of phonons in the BEC are identified; direct driving and parametric driving. Additionally, the oscillating gravitational field couples phonon modes, which can lead to the transition of excitations between modes. The special case of the gravitational field of a small, oscillating sphere located closely to the BEC is considered. It is shown that measurement of the effects may be possible for oscillating source masses down to the milligram scale, with a signal to noise ratio of the order of 10. To this end, noise terms and variations of experimental parameters are discussed and generic experimental parameters are given for specific atom species. The results of this article suggest the utility of BECs as sensors for the gravitational field of very small oscillating objects which may help to pave the way towards gravity experiments with masses in the quantum regime. 
\\ 

\noindent{\it Keywords\/}: Bose-Einstein condensation, phonons, oscillating gravitational fields

\end{abstract}
\pacs{04.80.Nn, 67.85.Hj, 67.85.Jk}

\maketitle

\section{Introduction}
\label{sec:introduction}

BECs are very small and extremely cold systems of a large number of atoms. These properties are famously exploited for high precision measurements of forces using atom interferometry \cite{Muentinga:2013int,Altin:2013pre,Tino:2014atom,Abend:2016atom,Hardman:2016sim,Asenbaum:2017pha}. Another method that utilizes BECs as sensors for forces is to measure the forces' effect on the collective oscillations of the atoms in the BEC. One specific example is the measurement of the thermal Casimir-Polder force \cite{Obrecht:2007mea}, which was theoretically proposed in \cite{Antezza:2004eff}. Further theoretical proposals to use collective oscillations in BECs and, in particular, their phonon modes for sensing purposes include, for example, gravitational wave detectors \cite{Sabin:2014bua,Sabin:2016ther,Howl:2018qua} , sensors for the effect of spacetime curvature on entanglement \cite{Bruschi:2014tes} (for a review see \cite{Howl:2016ryt}) and magnetic field and rotation sensors employing solitons formed by optical lattices \cite{Veretenov:2007int}. 

In this article, we investigate the effect of an oscillating gravitational field on the phonon modes of a BEC in a uniform trapping potential for the particular case of the gravitational near field of a small, oscillating gold or tungsten source mass. In particular, we show that the effect may be detectable for source masses down to the milligram scale  considering state of the art experimental parameters. The abilities of experimentalists to cool and control BECs of large numbers of atoms are advancing quickly and BECs may become sensors for very weak, oscillating gravitational fields in the future. The particular situation of a small source mass could be used to measure the gravitational field of a macroscopic quantum system in a spatial superposition state. The preparation of such a state is proposed in \cite{Romero-Isart:2011lar} and a proof of principle experiment for the measurement of the gravitational field of a small, oscillating mass was proposed in \cite{Schmole:2016mde}, which is based on a macroscopic test mass instead of a BEC. Furthermore, the experimental situation considered in this article may be useful to approach the experimental realization of the proposal to use the phonon modes of a BEC to detect gravitational waves presented in \cite{Sabin:2014bua} and the proposal to measure gravitationally induced Josephson tunneling presented in \cite{Sorge:2010qua}. While a relativistic framework is used in \cite{Sabin:2014bua} and \cite{Sorge:2010qua}, we consider only the effect of a Newtonian potential in this article. However, in \cite{Raetzel:2017zrl}, it was shown that the effect of a gravitational wave on the detector proposed in \cite{Sabin:2014bua} can be completely described in the Newtonian limit. 
Additionally, a Newtonian potential can be derived from a spacetime metric using the proper detector frame \cite{Ni:1978proper} in a similar way to the investigations in \cite{Raetzel:2017etl} for an optical resonator in curved spacetime. Furthermore, the equations describing the BEC in this article can be derived from the non-relativistic limit of the relativistic equations used for the derivations of the results in \cite{Sabin:2014bua} and \cite{Sorge:2010qua}. Therefore, the results presented in this article can be used to describe the measurement of gravitational effects with BECs in a general relativistic context using the techniques employed in \cite{Raetzel:2017etl}.

Besides the above mentioned ambitious benefits of an experimental realization of the situation presented in this article, such a realization would be the first investigation of the interaction of an oscillating gravitational field with the phonon modes of a BEC, which may be considered an achievement by itself.

We start our considerations by discussing our approximate description of the gravitational potential seen by the BEC in Sec. \ref{sec:gravfield}. In Sec. \ref{sec:descriptionfluid}, we introduce our first description of the BEC in terms of the order parameter and we introduce the basic equations for perturbations of the order parameter which represent the phonon modes. Then, we introduce dissipation in the phonon modes in Sec. \ref{sec:damping}, and we connect the amplitude of the perturbation of the order parameter to the number of created phonons in Sec. \ref{sec:phononenergy}. In Sec. \ref{sec:resdriv}, we investigate the effect of resonant driving by the gravitational near field of the source mass on the BEC. In Sec. \ref{sec:coherent}, the quantum field theoretical description of phonons is introduced and we derive the amplitudes for particle creation due to the oscillating sphere. In Sec. \ref{sec:exppar}, we derive expressions for experimental parameters required to achieve a given signal to noise ratio. Furthermore, we give a list of generic experimental parameters that may allow a detection of the phonons created by the gravitational acceleration and the gravity gradient, respectively, due to an oscillating massive sphere of mass $M$ for the two cases of $M=200\,\rm{g}$ and $M=0.2\,\rm{g}$. In Sec. \ref{sec:conclusions}, we conclude and discuss the possibility to use phonons in a BEC to detect oscillating gravitational fields and the influence of seismic and Newtonian noise on the noise background.

\section{The gravitational field of an oscillating sphere}
\label{sec:gravfield}

\begin{figure}[h]
\includegraphics[width=8cm,angle=0]{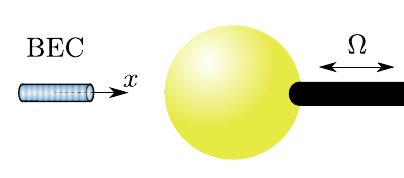}
\caption{\label{fig:becinfrontofgoldmass} The BEC is placed in front of a small massive sphere that oscillates at a lever with frequency $\Omega/2\pi$. The gravitational field of the sphere induces oscillations of the density of the BEC.}
\end{figure}

We assume that the BEC is of length $L$ in the oscillation direction of the sphere (see Fig. \ref{fig:becinfrontofgoldmass}) and the trap potential is a uniform box, as was realized in experiments such as the one described in \cite{Gaunt:2013bose}. Furthermore, we assume that the BEC is much smaller than the source mass, so that we can restrict our considerations to one spatial dimension.
Let us denote the distance from the center of the trap potential in the direction of the massive sphere as $x$ and let $R(t)$ be the distance between the center of the trap potential and the center of the sphere. For small $L$, the Newtonian potential of the moving sphere can be approximated, for all positions $x$ inside the trap potential, as
\begin{eqnarray}\label{eq:Newpotapprox}
	\Phi(t,x) & \approx & \Phi_0(t) - a(t)x - \mathfrak{G}(t)\frac{x^2}{2}\,,
\end{eqnarray}
where $\Phi_0(t)=-MG/R(t)$, $a(t)=MG/R(t)^2$ and $\mathfrak{G}(t)=2MG/R(t)^3$ are, respectively, the value of the gravitational potential at the center of the box trap potential, the gravitational acceleration due to the sphere evaluated at the center of the trap potential\footnote{Note that the source mass is placed in the positive $x$-direction from the BEC, which leads to an acceleration in the positive $x$-direction.} and its first derivative in the $x$-direction, the gravity gradient. By comparing the second and third term in Eq. (\ref{eq:Newpotapprox}) in the range of the BEC, we find that the second term is always larger than the third term by a factor $2R(t)/L$. Therefore, $\mathfrak{G}(t)$ becomes significant only for distances from the center of the sphere that are a few orders larger than the extension of the BEC trapping potential. When we investigate the perturbations of the BEC induced by the Newtonian potential (\ref{eq:Newpotapprox}), we will find that the gravity gradient leads to a very different signature than the acceleration. This is because the second term in Eq. (\ref{eq:Newpotapprox}) is linear in $x$, while the third term is quadratic in $x$. 

Let us assume that the source mass moves sinusoidally about a fixed position so that $R(t)=R_0 + \delta_R \sin(\Omega t+\varphi)$. If we assume that $R_0\gg \delta_R$, we can expand $a(t)$ and $\mathfrak{G}(t)$ in $\delta_R$. In the following, we will assume that $\delta_R/R_0$ is small enough to stop the expansion after the linear order in $\delta R$. We find
\begin{eqnarray}\label{eq:aRexpansion}
	 a(t) &\approx & \frac{MG}{R_0^2} \left(1-\frac{2\delta_R}{R_0}\sin(\Omega t+\varphi)\right)\quad\rm{and}\quad \mathfrak{G}(t) \approx  \frac{2MG}{R_0^3} \left(1 - \frac{3\delta_R}{R_0}\sin(\Omega t+\varphi)\right)\,.
\end{eqnarray}
 In an experiment it may be possible to move the mass such that the acceleration or the gravity gradient become exactly sinusoidally time dependent. Hence, the assumption of small values of $\delta_R/R_0$ is a restriction that may be lifted by a different ansatz and more detailed calculations. 

For $L=200\,\rm{\mu m}$ and $\delta_R=2\,\rm{mm}$, a minimal distance between the end of the box potential and the surface of the source mass $R_\rm{min}=1\,\rm{mm}$ and a $200\,\rm{g}$ tungsten or gold sphere, we have $R_0 = r + R_\rm{min} + L/2 + \delta_R \approx 17\,\rm{mm}$, where $r\approx 14\,\rm{mm}$ is the radius of the sphere. We find that the amplitudes of the terms proportional to $\sin(\Omega t+\varphi)$ in (\ref{eq:aRexpansion}) are 
\begin{equation}\label{eq:acccurvvalues}
	a_\Omega := \frac{2MG\delta_R}{R_0^3}\sim 2\times 10^{-8}\,\rm{ms^{-2}}\,\quad\rm{and}\quad \mathfrak{G}_{\Omega} := \frac{6MG\delta_R}{R_0^4} \sim 2\times 10^{-6}\,\rm{s^{-2}}\,.
\end{equation}
In the next section, we will introduce the description of the BEC and show how to derive the effect of the oscillating terms in the gravitational potential.

\section{BEC mean field perturbations}
\label{sec:descriptionfluid}

 One way to describe a BEC and the perturbations induced by an external potential is in terms of the order parameter $\psi$ governed by the Gross-Pitaevskii (GP) equation and perturbations governed by the Bogoliubov-DeGennes (BDG) equations derived from the GP equation. A second option is a microscopic description in terms of the atom field which is split into the condensate fraction and a phonon field. The condensate fraction is again described by the Gross-Pitaevskii equation. The phonon field evolution can be given by an interaction Hamiltonian and a mode decomposition with modes governed by the BDG equations. In Sec. \ref{sec:descriptionfluid} to Sec. \ref{sec:phononenergy}, we follow the first option. The second option is followed in Sec. \ref{sec:coherent}. Both approaches are only valid for BECs at low temperature. In particular, we need $k_BT\ll \mu$, where $\mu$ is the chemical potential of the condensate, $T$ its temperature (see Sec. \ref{sec:exppar} for explicit values of these parameters) and $k_B$ the Boltzmann constant. For $k_BT\ll \mu$, only a small part of the atoms is not in the ground state and its back action on the order parameter can be neglected for the purposes of this article\footnote{The effect of the thermal cloud of atoms can be taken into account by a generalized GP equation using the Popov approximation. For more details, see for example \cite{Giorgini:1997sca} and Sec. 13 of \cite{Pitaevskii:2003bose}. In particular, the back action of the thermal cloud leads to a stationary deformation of the order parameter, which would result in a perturbation of the number of created phonons. Since the number of atoms in the thermal cloud is very small in comparison to the number of atoms in the condensate fraction, the effect will be only a small change of the number of created phonons which we will neglect.}.
In this article, we include effects of finite temperature by using an effective description in terms of dissipation in the time evolution of phonons in Sec. \ref{sec:damping}. Furthermore, we consider effects of evaporating atoms in  Sec. \ref{sec:exppar}.

Since the gravitational potential (\ref{eq:Newpotapprox}) only depends on $x$, we only consider modulations of the order parameter in the $x$-direction and use an effectively one dimensional description. Then, the Gross-Pitaevskii (GP) equation is given as \cite{Pitaevskii:2003bose}
\begin{equation}\label{eq:gross}
	i\hbar \partial_t \psi = \left[-\frac{\hbar^2}{2m}\partial_x^2 + V +  \frac{\hbar^2}{2m} \lambda|\psi|^2 \right]\psi\,,
\end{equation}
where $m$ is the mass of the BEC atoms, $\lambda=8\pi a_\rm{scatt}$ describes the interaction between the atoms, $a_\rm{scatt}$ is the scattering length and $V=V_\rm{trap} + m\Phi$ is the external potential consisting of the trap potential $V_\rm{trap}$ and the Newtonian potential (\ref{eq:Newpotapprox}). To gain some intuition for the meaning of $\psi$, we can use the Madelung representation of the order parameter $\psi=\sqrt{\rho} e^{i\theta}$, where $\rho$ can be identified as the atom number density of the condensate and $\rho\hbar\partial_x\theta/m$ gives the probability current in the $x$-direction (see Eq. 5.10 of \cite{Pitaevskii:2003bose}), which can be interpreted as the spatial flow of atoms, while $\partial_t\theta$ is proportional to the chemical potential $\mu$ of the condensate in the case of a stationary solution. 

The effect of the gravitational potential of the source mass can be expected to be small. Therefore, it will lead to small perturbations of the atomic cloud. The equations that govern these perturbations can be derived from the GP equation \cite{Pethick:2002bose,Pitaevskii:2003bose}. To this end, we set $\psi=\sqrt{\rho_0} e^{i\theta_0 + i\phi_0}(1+\delta\psi)$, where we assume that $\sqrt{\rho_0}e^{i\theta_0}$ is a solution of the GP equation (\ref{eq:gross}) for the case where the gravitational potential vanishes and the density is stationary $\partial_t\rho_0=0$.  The complex function $\delta\psi$ is a space and time dependent perturbation that corresponds to phonons and $\phi_0$ is a spatially constant, time dependent function that captures the perturbation of the ground state energy. We rewrite the GP equation (\ref{eq:gross}) for the unperturbed solution $\sqrt{\rho_0} e^{i\theta_0}$ as the pair of coupled partial differential equations
\begin{eqnarray}\label{eq:madelung}
	\dot\theta_0\sqrt{\rho_0} + \frac{V}{\hbar}\sqrt{\rho_0}  + \frac{\hbar}{2m}\left(\theta_0'^2  + \lambda \rho_0\right)\sqrt{\rho_0} -\frac{\hbar}{2m}\nonumber \sqrt{\rho_0}'' &=&0\\
	2\sqrt{\rho_0}'\theta_0' + \sqrt{\rho_0}\theta_0'' &=&0\,,
\end{eqnarray}
where the prime and the dot denote the derivative in the $x$-direction and the time derivative, respectively. 

We consider a uniform box potential with infinite potential walls, which implies the boundary condition $\sqrt{\rho_0}=0$ at the potential walls. We assume that the length of the condensate $L$ is much larger then its healing length $\zeta = 1/\sqrt{\lambda \rho_0}$. Then, the density of the BEC can be assumed to be constant everywhere except for a small region close to the potential walls \cite{Pitaevskii:2003bose,Gaunt:2013bose,Gotlibovych:2014deg}. We can assume that $\rho_0$ is constant over the whole length of the uniform trap by additionally restricting our considerations to perturbations $\delta\psi$ with wavelength much larger than the healing length. In particular, this means that we do not consider quasi-particles that behave like free particles.

Inside the box potential, it follows from (\ref{eq:madelung}) and $\rho_0'=0$ that $\theta_0''=0$. At the boundary $\rho_0$ vanishes, $\rho_0'$ becomes singular and $\theta_0'$ has to vanish for the second equation in (\ref{eq:madelung}) to be fulfilled. Since $\theta_0''=0$ inside the box potential, it follows that $\theta_0'$ vanishes everywhere. We set $V_\rm{trap}=0$ for the region between the potential walls and we obtain the only remaining equation
\begin{equation}\label{eq:theta0dot}
	\dot\theta_0=-\frac{\mu}{\hbar}=-\frac{m}{\hbar}c_0^2 \,,
\end{equation}
where $\mu$ is the chemical potential of the BEC and $c_0$, the speed of sound in the BEC, is defined as
\begin{equation}\label{eq:soundspeed}
	c_0:=\sqrt{\frac{\hbar^2}{2m^2}\lambda \rho_0}=\frac{\hbar}{\sqrt{2}m\zeta}\,.
\end{equation}
Inserting the expansion  $\psi=\sqrt{\rho_0} e^{i\theta_0+i\phi_0}(1+\delta\psi)$ into Eq. (\ref{eq:gross}), and using Eq. (\ref{eq:theta0dot}) and all the properties of $\sqrt{\rho_0}$ and $\theta_0$ discussed above, we obtain the time dependent Bogoliubov-DeGennes equations in first order in the perturbation $\delta\psi$ and its complex conjugate \footnote{Note, if we set $\Phi=0$ and consider only points inside the trap, these equations lead to the time independent Bogoliubov-DeGennes equations for $\delta\psi = e^{-i\mu/\hbar}(u e^{-i\omega t} - v^* e^{i\omega t})$ that can be found in \cite{Pethick:2002bose} and other standard books.}
\begin{eqnarray}
	\label{eq:BDG1}  i\partial_t \delta\psi &=& -\frac{\hbar}{2m}\left(\frac{\rho_0^\prime}{\rho_0} \partial_x + \partial_x^2\right)\delta\psi + \frac{\delta V}{\hbar}\left(1+\delta\psi\right) + \frac{\hbar}{2m}\lambda\rho_0\left(\delta\psi + \delta\psi^*\right) \\
	\label{eq:BDG2}  - i\partial_t \delta\psi^* &=& -\frac{\hbar}{2m}\left(\frac{\rho_0^\prime}{\rho_0} \partial_x + \partial_x^2\right)\delta\psi^* + \frac{\delta V}{\hbar}\left(1+\delta\psi^*\right) + \frac{\hbar}{2m}\lambda\rho_0\left(\delta\psi + \delta\psi^*\right) \,,
\end{eqnarray}
where we defined $\delta V= \left(m\Phi - \delta\mu\right)$ and $\delta\mu := -\hbar\dot \phi_0$. Similar to the expansion of $\psi$, we can give an expansion of density and phase as $\sqrt{\rho}=\sqrt{\rho_0}(1+\alpha)$ and $\theta=\theta_0+\phi_0+\phi$. Via the equation $\psi=\sqrt{\rho} e^{i\theta}$, we can identify $\delta\psi$ and its complex conjugate with perturbations of density and phase as $\alpha = (\delta\psi + \delta\psi^*)/2$ and $\phi = -i(\delta\psi + \delta\psi^*)/2$. From Eqs. (\ref{eq:BDG1}) and (\ref{eq:BDG2}), we obtain
\begin{eqnarray}
	\label{eq:bulkpert1}  \dot\phi + \frac{\delta V}{\hbar}(1+\alpha)  + \frac{\hbar}{m \zeta^2} \alpha  - \frac{\hbar}{2m} \left(\frac{\rho_0'}{\rho_0}\alpha' + \alpha''\right)  &=&0\\
	\label{eq:bulkpert2}  \dot \alpha - \frac{\delta V}{\hbar}\phi + \frac{\hbar}{2m} \left(\frac{\rho_0'}{\rho_0}\phi' + \phi''\right) &=&0\,,
\end{eqnarray}
In Sec. \ref{sec:gravfield}, we found that the gravitational potential can be split into a stationary term and oscillating terms. Here, we are only interested in the oscillating terms which lead to oscillating perturbations $\alpha$ and $\phi$. The stationary term leads to a stationary perturbation $\alpha^\rm{c}$ and $\phi^\rm{c}$, which slightly changes the evolution equations for $\alpha$ and $\phi$. This leads to effects of higher order in the gravitational potential. In Appendix \ref{sec:stationary}, $\alpha^\rm{c}$ and $\phi^\rm{c}$ are derived and evaluated for appropriate experimental parameters. In the following, they will be neglected and we will consider $\delta V= \left(m\delta\Phi - \delta\mu\right)$, where $\delta\Phi=(\Phi_{0,\Omega} + a_{\Omega} x + \mathfrak{G} x^2/2)\sin(\Omega t + \varphi)$ and $\Phi_{0,\Omega}=MG\delta_R/R_0^2$.

The first part of the last term in Eqs. (\ref{eq:bulkpert1}) and (\ref{eq:bulkpert2}) vanishes inside the trap and becomes singular at the potential walls. Therefore, these terms can be neglected inside the trap and lead to the von Neumann boundary conditions $\alpha'=0=\phi'$ at the potential walls. We take the boundary conditions into account by using a mode decomposition 
\begin{eqnarray}\label{eq:expansion}
	\phi &=& \sum_{n\ge 1} g_n(t)\varphi_n(x)\quad\rm{and}\quad \alpha = \sum_{n\ge 1} f_n(t)\varphi_n(x)\,,
\end{eqnarray}
where $\varphi_n(x)=\cos(k_n(x+L/2))$ and $k_n=n\pi/L$. Note that the total number of atoms in the condensate given by $N:=\mathcal{A}\int_{-L/2}^{L/2} dx \,\rho_0(1+\alpha)^2$ is a constant in first order in the perturbation, where $\mathcal{A}$ is the cross sectional area of the BEC in the $y$-$z$-plane. 

Projecting Eqs. (\ref{eq:bulkpert1}) and (\ref{eq:bulkpert2}) on the mode $\varphi_n(x)$ as $(\varphi_n,\alpha)=\int_{-L/2}^{L/2} dx\, \varphi_n \alpha$, we find for each $n$ the equations
\begin{eqnarray}
	\label{eq:bulkpertproj1}  \dot g_n + \frac{2}{L}\int_{-L/2}^{L/2} dx\,\frac{\delta V}{\hbar}(1+\alpha)\varphi_n  + \frac{\hbar}{m \zeta^2} f_n  + \frac{\hbar}{2m} k_n^2 f_n  &=&0\\
	\label{eq:bulkpertproj2}  \dot f_n - \frac{2}{L}\int_{-L/2}^{L/2} dx\, \frac{\delta V}{\hbar}\phi\,\varphi_n - \frac{\hbar}{2m}k_n^2 g_n &=&0\,,
\end{eqnarray}
As mentioned above, we only consider perturbations with wavelengths much larger than the healing length, which can be expressed as $k_n\zeta \ll 1$. Therefore, the last term in Eq. (\ref{eq:bulkpertproj1}) is much smaller that the second to last term and can be neglected. 
Then, by multiplying Eq. (\ref{eq:bulkpertproj1}) by $\dot{\delta V}$, we obtain that $\dot{\delta V}f_n = -m\zeta^2\dot{\delta V}\dot g_n/\hbar$ in first order in the potential perturbation. Taking this into account, taking the first time derivative of Eq. (\ref{eq:bulkpertproj1}) and using Eq. (\ref{eq:bulkpertproj2}) to replace $\dot f_n$, we obtain for each $n$
\begin{eqnarray}\label{eq:projected}
	\ddot g_n  + \frac{2}{L}\int_{-L/2}^{L/2} dx\,\sum_l\left[-\frac{m\zeta^2\dot{\delta V}}{\hbar^2}\dot g_l  + \frac{\delta V}{m\zeta^2}g_l\right]\varphi_l\varphi_n + \omega_n^2 g_n  &=& -\frac{2}{L}\int_{-L/2}^{L/2} dx\,\frac{\dot{\delta V}}{\hbar}\varphi_n
\end{eqnarray}
where we approximated $(1 - \zeta^2k_l^2)g_l \approx g_l$ because we assumed $k_l\zeta \ll 1$, and where we defined $\omega_n = c_0 k_n$. This is a set of coupled, driven harmonic oscillators, which for vanishing driving, evolve with frequency $\omega_n$. Let us assume that the frequency of the external driving $\Omega$ is of the same order as $\omega_n$ such that $|\dot{\delta V}| \sim \omega_n | \delta V|$. Then, we obtain that the absolute value of the first term in the integral in Eq. (\ref{eq:projected}) is proportional to $m\omega_n\omega_l\zeta^2\hbar^{-2} |\delta V g_l| = k_n k_l(2m)^{-1} |\delta V g_l|$, while the second term is proportional to $(m\zeta^2)^{-1}\delta V g_l$. In this article, we will only consider situations in which just one mode $l$ contributes to the evolution of the mode $n$ via the coupling term in the integral in Eq. (\ref{eq:projected}).\footnote{The inevitable contribution of all modes via thermal excitations is taken into account via the damping term discussed in Sec. \ref{sec:damping}.} Then, since $k_n\zeta \ll 1$ and $k_l\zeta \ll 1$, the second term in the integral in Eq. (\ref{eq:projected}) will dominate significantly and we can neglect the first term. Finally, we find a set of ordinary, coupled linear differential equations for the amplitudes $g_n$ that are driven by the gravitational field
\begin{equation}\label{eq:harmosc}
	 \ddot g_n + \omega_n^2(1+S_n)  g_n = D_n + \sum_{l\neq n} T_{nl} g_l\,,
\end{equation}
where
\begin{eqnarray}\label{eq:momentsall}
	\nonumber D_n := -\frac{2}{\hbar L} \int_{-L/2}^{L/2}dx\, \dot{\delta V}\varphi_n\,,\quad S_n := \frac{2}{\omega_n^2 m\zeta^2 L} \int_{-L/2}^{L/2}dx\, \delta V \varphi_n^2\quad\rm{and}\quad
	\nonumber T_{nl} := -\frac{2}{m\zeta^2 L} \int_{-L/2}^{L/2}dx\, \delta V \varphi_n\varphi_l\,.
\end{eqnarray}
Eq. (\ref{eq:harmosc}) is the evolution equation for a driven harmonic oscillator with three different driving mechanisms; there is direct driving through $D_n$, the driving due to other excited modes through $T_{nl}$ and parametric driving through $S_n$.
Finally, we have to specify $\delta\mu=-\hbar\dot\phi_0$. For that purpose, we project Eq. (\ref{eq:bulkpert1}) on the constant function representing the density distribution of the ground state. It follows that $\delta\mu = -\hbar\dot \phi_0 = m\int_{-L/2}^{L/2} dx \,\delta\Phi/L$, which is the spatial average of the Newtonian potential.
Then, we obtain for the driving moments, the parametric frequency modulation and the mode coupling coefficients
\begin{eqnarray}\label{eq:momentsall2}
	\nonumber D_n &=& \left(\left(1-(-1)^n\right)\frac{a_{\Omega}}{L} - \left(1+(-1)^n\right) \frac{\mathfrak{G}_{\Omega}}{2}\right)\frac{2L^2 m\Omega}{n^2\pi^2\hbar}\cos(\Omega t+\varphi)\,,\quad  S_n =  \frac{ \mathfrak{G}_{\Omega} m^2L^4}{2n^4\pi^4\hbar^2} \sin(\Omega t+\varphi)\quad\rm{and}\\
	 T_{nl} &=& \left(\left(1-(-1)^{l+n}\right)\frac{a_{\Omega}}{L}  - \left(1+(-1)^{l+n}\right) \frac{\mathfrak{G}_{\Omega}}{2}\right) \frac{2L^2(l^2+n^2)}{\zeta^2(l^2-n^2)^2\pi^2}\sin(\Omega t+\varphi)\quad\rm{for}\quad l\neq n\,,
\end{eqnarray}
respectively.  We see the different signatures of the acceleration and the gravity gradient in the perturbation of the BEC. Via the direct driving $D_n$, the gravity gradient couples to the symmetric modes while the acceleration couples to the anti-symmetric modes in the range of validity of our approximations. Similarly, acceleration couples modes of different parity and the gravity gradient couples modes of the same parity. Parametric driving $S_n$ affects all modes, but it is only due to the gravity gradient. 

To obtain the full evolution of the phonon modes in the BEC, we have to include all significant damping effects. This is the content of the next section.

%%%%%%%%%%%%%%%%%%%%%%%%%%%%%%%%%%%%%%%%%%%%%%%%%

\section{Damping}
\label{sec:damping}

A phenomenological way to describe damping on the level of the Gross-Pitaevskii equation was introduced in \cite{Pitaevskii:1959phen} and discussed further in \cite{Choi:1998ia}. We discuss this approach in Appendix \ref{sec:dissipation}. It is equivalent to adding a damping term $\gamma_n \dot g_n$ to the left hand side of the harmonic oscillator Eq. (\ref{eq:harmosc}), which leads to
\begin{equation}\label{eq:harmoscdamp2}
	 \ddot g_n + \gamma_n \dot g_n + \omega_n^2(1+S_n)g_n = D_n + \sum_{l\neq n} T_{nl} g_l\,,
\end{equation}
In the following, we will discuss the two most important mechanisms of damping and their contribution to the damping constant $\gamma_n$.  There is a great deal of literature about damping of phonon modes in uniform BECs with periodic boundary conditions. In Appendix \ref{sec:dampbox}, we show that damping of phonon modes in a box potential with von Neumann boundary conditions can be described approximately using the expressions for damping in a uniform BEC with periodic boundary conditions. We will only discuss these expressions in the following.

For BECs at temperatures $T$ such that $k_B T\gg \hbar \omega_n$, where $k_B$ is the Boltzmann constant, one mechanism of damping is Landau damping, in which energy from the perturbations of the mean field is absorbed by the thermal bath of excitations. Landau damping in BECs was initially discussed in \cite{Szepfalusy:1974on,Shi:1998fin,Fedichev:1998damp}. An expression for the damping constant $\gamma_n$ in a uniform BEC for general temperatures was derived in \cite{Pitaevskii:1997land,Giorgini:1998dam} and it is given as
\begin{equation}\label{eq:Landaudamp}
 	\gamma_n^\rm{La} = \frac{mc_0 \lambda \omega_n }{2\pi \hbar} \int_0^\infty dx\,(e^x-e^{-x})^{-2}\left(1-\frac{1}{2u}-\frac{1}{2u^2}\right)^2\,,
\end{equation}
where $u=\sqrt{1+4(k_BT/\mu)^2 x^2}$. For temperatures such that $k_BT \gg \mu $, we obtain the expression
\begin{equation}\label{eq:Landaudamphigh}
 	\gamma_n^\rm{La} = \frac{3}{64}\frac{\lambda k_B T \omega_n}{\hbar c_0}\,,
\end{equation}
which delivers values that are in agreement with experiments like \cite{Jin:1997temp} (see for example \cite{Dalfovo:1999theo} for a discussion). For temperatures such that $k_BT \ll \mu $ the Landau damping rate becomes
\begin{equation}\label{eq:Landaudamplow}
 	\gamma_n^\rm{La} = \frac{3\pi^3}{40}\frac{(k_B T)^4 \omega_n}{m \rho_0 \hbar^3 c_0^5}\,,
\end{equation}
Note that (\ref{eq:Landaudamphigh}) is linear in the temperature, while (\ref{eq:Landaudamplow}) is proportional to its fourth power, which means that the Landau damping can be lowered significantly by lowering the temperature further once the low temperature regime is reached. For the experimental parameters that we will consider in Sec. \ref{sec:exppar} we have $k_BT \ll \mu $ and Landau damping is described by Eq. (\ref{eq:Landaudamplow}).

Another contribution to the damping rate that becomes most significant for low temperatures arises through Beliaev damping. The microscopic origin of Beliaev damping is the decay of a single phonon into two phonons of lower energy. The corresponding damping constant is given as \cite{Giorgini:1998dam}
\begin{equation}\label{eq:Beliavdamp}
	\gamma_n^\rm{Be} = \gamma_n^\rm{Be,0} \left[1 + 60 \int_0^1dx\frac{x^2(x-1)^2}{e^{\frac{xc\hbar k_n}{k_BT}}-1}\right]\,,
\end{equation}
where
\begin{equation}\label{eq:Beliavdamplow}
	\gamma_n^\rm{Be,0} = \frac{3}{640\pi}\frac{\hbar  k_n^5}{m  \rho_0}\,
\end{equation} 
is the Beliaev damping constant at zero temperature, and we assumed that the quotient of the atom density of the BEC and total atom density including the thermal cloud is close to one, which restricts our considerations to temperatures much smaller than the critical temperature of the BEC \cite{Giorgini:1997sca}. Note that (\ref{eq:Beliavdamplow}) is proportional to the fifth power of the wave number while the Landau damping constant is linear in the wave number. This means that Beliaev damping becomes dominant for higher wave numbers. Since Beliaev damping remains even for zero temperature, we can conclude that the value of the wave number at which Beliaev damping becomes dominant is lower for lower temperatures. 

The third mechanism of damping that has to be taken into account is damping due to atomic losses \cite{Dziarmaga:2003bose}; the atom-atom interactions and thermal fluctuations of the Bose gas lead to the evaporation of atoms from the condensate. The atom loss leads to damping and fluctuations of the phonon modes. In \cite{Dziarmaga:2003bose}, it was shown that the numerical value of the corresponding damping rate $\gamma^\rm{loss}$ is approximately the same as the evaporation rate of the condensate independently of the mode number. In general, the total damping rate can be written as
\begin{equation}\label{eq:gammatotal}
 	\gamma_n = \gamma_n^\rm{La} + \gamma_n^\rm{Be} + \gamma^\rm{loss}\,.
\end{equation}
In the next section, we will discuss the number of phonons that correspond to the amplitude $g_n$.

%%%%%%%%%%%%%%%%%%%%%%%%%%%%%%%%%%%%%%%%%%%%%%%%%

\section{Average number of phonons in the mean field perturbations}
\label{sec:phononenergy}

The picture of perturbations of the order parameter $\psi$ hides the true nature of the perturbations as phonons. 
However, the amplitude $g_n$ of the phase perturbation corresponds to a certain number of phonons $\bar N_{n,\rm{c}}$ that we can obtain by deriving the amplitude of the phase perturbation corresponding to a single phonon.

For this purpose, we consider the energy of free fluctuations of the BEC density and phase without driving, which is given as
\begin{eqnarray}\label{eq:energyfunctional}
	\Delta\mathcal{E}[\alpha,\phi] &=& \frac{\hbar^2}{2m}\mathcal{A}\int_{-L/2}^{L/2} dx \rho_0\, \left( 2\zeta^{-2}\alpha^2  - \phi\phi'' - \alpha \alpha'' \right)	\,,
\end{eqnarray}
where we used Eq. (5.76) of \cite{Pitaevskii:2003bose}, the replacement $\vartheta=\sqrt{\rho_0}(\alpha+i\phi)$ and the equations of motion (\ref{eq:bulkpert1}) and (\ref{eq:bulkpert2}) for $\delta V=0$ and considering the von Neumann boundary conditions to be fulfilled. For the modes defined in Eq. (\ref{eq:expansion}), we obtain the energy 
\begin{eqnarray}
	\Delta\mathcal{E}_n	&=& \frac{\hbar^2k_n^2 N_a}{4m}  \left( \frac{2}{\zeta^2 k_n^2}  f_n(t)^2 +   g_n(t)^2\right) 	\,,
\end{eqnarray}
where $N = V \rho_0 = \mathcal{A}L \rho_0$ is the number of atoms in the BEC and we neglected the contribution of the third term in Eq. (\ref{eq:energyfunctional}), since $\zeta k_n \ll 1$. Without driving, Eq. (\ref{eq:bulkpert1}) leads to $f_n = -m\zeta^2\dot g_n/\hbar$. Furthermore, a solution of the equations of motion (\ref{eq:harmosc}) without driving will be $g_n(t)=\bar g_{n}\sin(\omega_n t + \varphi_n)$ for some phase $\varphi_n$. The same result is found for the steady state solutions on resonance later in Eq. (\ref{eq:resonancedamped}). Therefore, using the linear dispersion $\omega_n=c_0 k_n$, we find 
\begin{eqnarray}\label{eq:energyampl}
	\Delta\mathcal{E}_n	&=& \frac{\hbar^2 k_n^2 N_a}{4m}   \bar g_{n}^2 \,.
\end{eqnarray}
The energy of a single phonon of mode $n$ is given as $\hbar\omega_n$ and we obtain the amplitude corresponding to a single phonon as $g_{n,\rm{ph}}=(2\sqrt{2}/(k_n \zeta N_a))^{1/2}$.
Additionally, we find for the average number of phonons in the coherent state created by the gravitational field of the moving mass  
\begin{equation}\label{eq:nofph}
	\bar N_{n,\rm{c}} = \frac{n\pi \zeta N_a}{2\sqrt{2}L} \bar g_{n}^2 \,.
\end{equation}
In Sec. \ref{sec:resdriv}, we will derive the amplitude of the dynamical modes of density perturbations induced by the moving mass and use Eq. (\ref{eq:nofph}) to extract the number of created phonons from the amplitude $\bar g_n$. From Eq. (\ref{eq:nofph}), we can also find a condition on the maximal number of phonons in a mode for which our approach of considering $\delta\psi$ as a small perturbation is still suitable. We find that $g_{n,0}\le 10^{-1}$ leads to $\bar N_{n,\rm{c}}\le 10^{-2}n\pi\zeta N_a/(2\sqrt{2}L)$.

%%%%%%%%%%%%%%%%%%%%%%%%%%%%%%%%%%%%%%%%%%%%%%%%%

\section{Phonon creation for resonant driving}
\label{sec:resdriv}

In Sec. \ref{sec:descriptionfluid} and \ref{sec:damping}, we derived the linearized differential equations for the evolution of the BEC under the gravitational influence of a moving source mass. In this section, we will derive the effect of the sinusoidal driving by solving Eq. (\ref{eq:harmoscdamp2}) for particular cases.

\subsection{Direct driving}

The first case that we want to consider is the case of direct driving, when the initial excitation of all modes can be considered to be zero. Then, Eq. (\ref{eq:harmoscdamp2}) reduces to 
\begin{equation}\label{eq:harmoscdampdirect}
	 \ddot g_n + \gamma_n \dot g_n + \omega_n^2 g_n = F_n \sin(\Omega t + \tilde\varphi)\,,
\end{equation}
where $\tilde\varphi=\varphi + \pi/2$ and
\begin{eqnarray}\label{eq:omegaFn}
	F_n&=&\left(\left(1-(-1)^n\right)\frac{a_{\Omega}}{L}  - \left(1+(-1)^n\right) \frac{\mathfrak{G}_{\Omega}}{2}\right)\frac{2L^2 m\Omega}{n^2\pi^2\hbar}\,.
\end{eqnarray}

The steady state solution on resonance for this driven and damped harmonic oscillator is
\begin{equation}\label{eq:resonancedamped}
	g_{\rm{st},n,\Omega\approx\omega_n}(t)=\frac{F_n }{\gamma_n\omega_n}\sin\left(\omega_n t  + \varphi_0 + \tilde\varphi \right)\,,
\end{equation}
where $\varphi_0=\pi/2$ if $\Omega$ is larger than $\omega_n$ and $\varphi_0=-\pi/2$
for $\Omega$ smaller than $\omega_n$. We see that the steady state amplitude is $F_n\omega_n/\gamma_n$. From the solution (\ref{eq:resonancedamped}), we see that there is a phase shift between the driving and the motion of the BEC of $\pi/2$ on resonance.

Instead, if we consider times much shorter than the inverse damping rate, the solution is the undamped solution
\begin{eqnarray}\label{eq:solutionfn}
	 g_{n,\gamma_n=0}(t) &=&  \frac{F_n}{\omega_n^2-\Omega^2}\left(\sin(\Omega t + \tilde\varphi) - \left(\sin\tilde\varphi \cos\left(\omega_n t \right) + \frac{\Omega}{\omega_n}\cos\tilde\varphi\sin\left(\omega_n t \right)  \right)\right)
\end{eqnarray}
which, for resonance becomes
\begin{eqnarray}\label{eq:solutionfnres}
		g_n(t) &=& - \frac{F_n}{2\omega_n^2}  \left(\omega_n t \cos(\omega_n t + \tilde\varphi) - \sin(\omega_n t)\cos(\tilde\varphi)\right)
\,.
\end{eqnarray}
Hence, the amplitude grows linearly in time for $t\ll \gamma_n^{-1}$. Together with Eq. (\ref{eq:solutionfnres}), Eq. (\ref{eq:nofph}) gives the number of coherent phonons created by the oscillating gravitational field of the massive sphere on resonance for times much shorter than the inverse damping rate of the mode under consideration as
\begin{equation}\label{eq:ncr}
	\bar N_{n,\rm{c},t\ll \gamma_n^{-1}} (t) \approx  \frac{m^2 \sqrt{2}\zeta N_a L^3 t^2}{\hbar^2 (n\pi)^3} \left(\frac{1-(-1)^n}{2} \left(\frac{a_{\Omega}}{L}\right)^2  + \frac{1+(-1)^n}{2} \left(\frac{\mathfrak{G}_{\Omega}}{2}\right)^2\right)
\end{equation}
From the values for acceleration and gravity gradient in Eq. (\ref{eq:acccurvvalues}) and Eq. (\ref{eq:ncr}), we see that much less phonons are created due to the oscillating gravity gradient than through the oscillating acceleration when only the direct driving term $D_n$ is considered.

\subsection{Mode coupling}

Of course, in general, the solutions (\ref{eq:solutionfnres}) and (\ref{eq:resonancedamped}) are only reliable if the inter-mode coupling can be neglected. From the expression for $T_{nl}$ in (\ref{eq:momentsall2}), we see that the driving term due to a coherent excitation of mode $l$ is of the same order as the direct driving term $D_n$ if
\begin{equation}
	\bar g_{l} \sim \bar g^\rm{lim,n}_{l}:=\frac{\pi \zeta}{\sqrt{2} L}\frac{(l^2-n^2)^2}{n(l^2+n^2)}	
	\quad\rm{or}\quad \bar N_l \sim \bar N^\rm{lim,n}_{l} :=  N_a \left(\frac{\pi \zeta }{\sqrt{2} L}\right)^3 \frac{(l^2-n^2)^4}{2n(l^2+n^2)^2}\,.
\end{equation}
If the inter-mode coupling cannot be neglected, either phonon pairs are created in modes $n$ and $l$, or phonons in an excited mode will be shifted into a mode of higher energy. This can be seen by considering $g_{l}$ to oscillate with frequency $\omega_l$ as $g_l = \bar g_{l}\sin(\omega_l t + \varphi_l)$. Neglecting parametric driving and direct driving, we find the differential equation
 \begin{equation}\label{eq:harmoscdampcoupling}
	 \ddot g_n + \gamma_n \dot g_n + \omega_n^2 g_n = 2G_{ln} \sin(\Omega t + \varphi)\sin(\omega_l t + \varphi_l)\,,
\end{equation}
where 
\begin{eqnarray}\label{eq:omegaGl}
	G_{ln}&=& \frac{\bar g_{l}}{2}\left(\left(1-(-1)^{l+n}\right)\frac{a_{\Omega}}{L}  - \left(1+(-1)^{l+n}\right) \frac{\mathfrak{G}_{\Omega}}{2}\right) \frac{2L^2(l^2+n^2)}{\zeta^2(l^2-n^2)^2\pi^2}\,.
\end{eqnarray}
The right hand side of Eq. (\ref{eq:harmoscdampcoupling}) can be rewritten as
\begin{equation}
	2G_{ln}\sin(\Omega t + \varphi)\sin(\omega_l t + \varphi_l) =G_{ln}[\cos((\Omega - \omega_l) t +  \varphi  - \varphi_l) - \cos((\Omega + \omega_l) t +  \varphi  + \varphi_l)]
\end{equation}
Therefore, resonant driving of mode $n$ is achieved if $\Omega=\omega_l+\omega_n$ or $\Omega = |\omega_n - \omega_l|$, which correspond to the creation of phonon pairs and the shift of phonons between modes, respectively. These processes will appear as multi-mode squeezing and mode mixing, respectively, in Sec. \ref{sec:coherent}. Assuming initial excitation of mode $n$ as $g_n(t)=g_{n,0}\sin(\omega_n t + \varphi_0)$, shorter times than the inverse damping constant and neglecting the back action on mode $l$, we find analogously to Eq. (\ref{eq:solutionfnres}) on resonance
\begin{eqnarray}\label{eq:couplingsol}
	g_n(t) &=& g_{n,0}\sin(\omega_n t + \varphi_0) - \frac{G_{ln}}{2\omega_n^2}  \left(\omega_n t \cos(\omega_n t + \tilde\varphi_l) - \sin(\omega_n t)\cos(\tilde\varphi_l)\right) \,,
\end{eqnarray}
 where $\tilde\varphi_l=\varphi-\varphi_l+\pi/2$ for $\Omega=\omega_n+\omega_l$ and  $\tilde\varphi_l=\varphi+\varphi_l - \pi/2$ for $\Omega=\omega_n-\omega_l$ and $\tilde\varphi_l=-\varphi+\varphi_l + \pi/2$ for $\Omega=\omega_l-\omega_n$. We see that, depending on the phase relation between driving and initial excitation of the mode $l$, the amplitude of the oscillation will increase from the start or it will decrease first and increase later. Hence, for an appropriately chosen phase relation, the mode coupling may be associated with a damping process. For values of $\bar N^\rm{lim,l}_n\ll 1$, such damping processes may be stronger than the direct driving $D_n$. In this situation, it may be more efficient when initial phonons are prepared in a certain mode and the gravitational field is measured through the induced loss of phonons from that mode. This possibility will be discussed in more detail in Sec. \ref{sec:coherent} and Sec. \ref{sec:exppar}. We have to keep in mind that we neglected the back action on mode $l$, and therefore, Eq. (\ref{eq:couplingsol}) can only hold for short time scales.

For initially vanishing excitation, $g_{n,0}=0$, and $\tilde\varphi_l=\pi/2$, we find for the number of created phonons in the mean field
\begin{equation}\label{eq:couplingcreated}
	\bar N_{n,\rm{c},t\ll \gamma_n^{-1}} = \bar N_{l,\rm{c},0}\left(\frac{ \sqrt{2} m t}{\hbar\zeta} \frac{L^3(l^2+n^2)}{\sqrt{nl}(l^2-n^2)^2\pi^3}\right)^2 \left(\frac{1-(-1)^{l+n}}{2} \left(\frac{a_{\Omega}}{L}\right)^2  + \frac{1+(-1)^{l+n}}{2} \left(\frac{\mathfrak{G}_{\Omega}}{2}\right)^2\right)
\end{equation}
where $\bar N_{l,\rm{c},0}=l\pi \zeta N_a \bar g_{l,0}^2/(2\sqrt{2}L)$ is the number of initial phonons in mode $l$ that contribute to the mean field.

\subsection{Parametric driving}

As the third process of interest in this system, let us investigate the effect of the parametric driving term $S_n=\bar S_n \sin(\Omega t + \varphi)$. For a classical system the parametric driving can only lead to an excitation of the system when the system is already excited. If we neglect all other processes and damping, start from an initial excitation that oscillates as $g_n(t)=\bar g_{n,0}\cos(\omega_n t + \varphi)$ for $t<0$ and consider parametric resonance $\Omega=2\omega_n$, the amplitude after a time $t$ is found as
\begin{equation}\label{eq:solpara}
	\bar g_n(t) \approx \bar g_{n,0} \exp\left( \frac{\bar S_n\omega_n t}{4} \right) = \bar g_{n,0} \exp\left( \frac{m \mathfrak{G}_\Omega L^3 t}{8\sqrt{2}n^3\pi^3\hbar \zeta} \right) 
\end{equation}
where $g_0$ is the initial amplitude. With Eq. (\ref{eq:nofph}), this leads to the expression for the average number of created phonons
\begin{equation}\label{eq:solparanumber}
	\delta\bar N_{n,\rm{c},t\ll \gamma_n^{-1}} (t)  =  \frac{n\pi \zeta N_a}{2\sqrt{2}L}(\bar g_n(t)^2-(\bar g_{n,0})^2)\approx \bar N_{n,0}\left(\exp\left( \frac{m \mathfrak{G}_\Omega L^3 t}{4\sqrt{2}n^3\pi^3\hbar \zeta} \right)-1\right)
\end{equation}
where $\bar N_{n,0}$ is the initial number of phonons in mode $n$ that contribute to the variation of the mean field.

In the next section, we will describe the phonons in the BEC as a quantum field and we will investigate the inter-mode coupling and the parametric driving in detail. We will recover Eq. (\ref{eq:ncr}) as the number of created coherent phonons, Eq. (\ref{eq:couplingcreated}) from mode mixing and Eq. (\ref{eq:solpara}) from single-mode squeezing of a coherent initial state.

\section{Quantum field description}
\label{sec:coherent}

For a description of the perturbations of the BEC as phonons, we start from a time dependent external potential\footnote{For other approaches to the description of BECs in time dependent potentials see for example \cite{Minguzzi:1997lin,Khawaja:2009sol,Damski:2001simp,Castin:1996bose,Yan:2012mat}} $V(\vec x,t) = V_\rm{trap}(\vec x) + \delta V_0(x) + \delta V(x,t)$, where  $\delta V(x,t)=\delta \bar{V}(x) \sin\Omega t$ and $\delta \bar{V} = m(\Phi_{0,\Omega} + a_{\Omega}x + \mathfrak{G} x^2/2)$. The external potential enters the Hamiltonian of the BEC as $\int dx^3\, V\hat\Psi^\dagger \hat\Psi$ such that the full Hamiltonian is given by (see \cite{Pitaevskii:2003bose} for the free Hamiltonian)
\begin{eqnarray}
	\hat H &=& \int_\mathcal{V} d^3x\, \hat\Psi^\dagger\left(-\frac{\hbar^2}{2m} \nabla^2  + V \right)\hat\Psi + \frac{g}{2}\int_\mathcal{V} d^3x  \,\hat\Psi^\dagger \hat\Psi^\dagger \hat\Psi \hat\Psi\,.
\end{eqnarray}
where $g=\hbar^2\lambda/(2m)$ and $\mathcal{V}$ is the volume of the box potential. We neglect the contribution of the stationary part $\delta V_0$ as in Sec. \ref{sec:descriptionfluid}, consider the box trap potential $V_\rm{trap}$ and the expansion \footnote{Here we are using the absolute perturbation $\hat\vartheta$ in contrast to the description in Sec. \ref{sec:descriptionfluid}.}
\begin{eqnarray}
	\hat\Psi(\vec x,t)= (\hat\Psi_0(\vec x) + \hat\vartheta(\vec x,t))e^{-i\mu t/\hbar - i\int_0^t dt'\,\delta\mu(t')/\hbar}
\end{eqnarray} 
in the Heisenberg picture, where $\hat\Psi_0 = \hat a_0 \bar\psi_0$, $\bar\psi_0 = (L\mathcal{A})^{-1/2}$ and $\delta\mu = \int_{-L/2}^{L/2} dx\, \delta V/L$ is the time dependent energy shift of the ground state. Furthermore, we apply the Bogoliubov approximation, which replaces $\hat a_0 \rightarrow \sqrt{N_a} \mathbb{I}$, throughout the interaction process. Then, we obtain the grand canonical interaction Hamiltonian (see Appendix \ref{sec:intham} for the detailed derivation)
\begin{eqnarray}\label{eq:Hprimeint}
	  \hat H_\rm{int} & = &  \sqrt{\rho_0} \int_\mathcal{V} d^3x\, (\delta V -\delta\mu)\left(\hat\vartheta + \hat\vartheta^{\dagger}\right) +  \int_\mathcal{V} d^3x\, \hat\vartheta^{\dagger}(\delta V -\delta\mu)\hat\vartheta\,.
\end{eqnarray}
The field operator is expanded as
\begin{equation}
	\hat\vartheta = \sum_n  \left(u_n\hat b_{n}e^{-i\omega_n t} + v_n\hat b_{n}^\dagger e^{i\omega_n t}\right)\,,
\end{equation}
where $[b_{n},b_{m}^\dagger]=\delta_{nm}$. The real mode functions fulfill the stationary Bogoliubov-deGennes (BDG) equations
\begin{eqnarray}
	 \omega_n u_n &=&\frac{\hbar}{2m}\left( \left(- \nabla^2 + \frac{1}{\zeta^2}\right)u_n + \frac{1}{\zeta^2} v_n\right)\\
	- \omega_n v_n &=& \frac{\hbar}{2m}\left(\left(- \nabla^2 + \frac{1}{\zeta^2}\right)v_n + \frac{1}{\zeta^2} u_n\right)\,,
\end{eqnarray}
fulfill von Neumann boundary conditions at the potential walls (due to the vanishing density $\rho_0$) and are normalized with respect to the inner product
\begin{equation}\label{eq:normauv}
	\int_\mathcal{V} d^3x \left(u_n u_m - v_n v_m\right) = \delta_{nm}\,.
\end{equation}
In the following, we restrict our considerations to modes with vanishing transversal wave numbers, i.e. we only consider the $x$-direction. With these considerations, we obtain the set of solutions of the BDG equations
\begin{eqnarray}\label{eq:modefunctionsuv}
	u_n = \alpha_n\varphi_n\quad\rm{and}\quad v_n = \beta_n\varphi_n\,
\end{eqnarray}
inside the box potential, where $\alpha_n = ((\sqrt{2}\zeta k_n)^{-1} + 1)^{1/2}(L\mathcal{A})^{-1/2}$ and $\beta_n = -((\sqrt{2}\zeta k_n)^{-1} - 1)^{1/2}(L\mathcal{A})^{-1/2}$. Furthermore $\varphi_n=\cos(k_n(x+L/2))$ and $k_n=\pi n/L$ were already defined before in Sec. \ref{sec:descriptionfluid}. This leads to $\omega_n=c_0 k_n$ for $k_n\zeta\ll 1$. Assuming that $\delta V(x,t)=\delta \bar{V}(x) \sin\Omega t$, $\delta\mu = \delta \bar\mu \sin\Omega t$ and the rotating wave approximation, the interaction Hamiltonian can be written in the interaction picture as
\begin{eqnarray}\label{eq:HIdeltaVmain}
	  \nonumber \hat H^{\delta V}_\rm{int} &=&  \sum_n M_{0n}\left(  \hat b_n e^{-i(\omega_n-\Omega) t} - \hat b^\dagger_n e^{i(\omega_n-\Omega)t}\right)  + \sum_{l,n} M_{ln}\left(\hat b_l \hat b_n e^{-i(\omega_n + \omega_l - \Omega) t} - \hat b_l^\dagger \hat b_n^\dagger e^{i(\omega_n + \omega_l - \Omega) t}\right) \\
	 && - \sum_{l > n}\left( A_{ln} \left(\hat b_l^\dagger \hat b_n e^{i(\omega_l - \omega_n - \Omega) t} - \hat b_n^\dagger \hat b_l e^{-i(\omega_l - \omega_n - \Omega) t} \right)  + B_{ln} \left(\hat b_l^\dagger\hat b_n  e^{i(\omega_l - \omega_n - \Omega) t} - \hat b_n^\dagger \hat b_l  e^{-i(\omega_l - \omega_n - \Omega) t} \right) \right)   \,,
\end{eqnarray}
where the transition amplitudes are given as
\begin{eqnarray}\label{eq:moments}
	M_{0n}&=&-\frac{i}{2}\sqrt{\frac{N_a\mathcal{A}}{L}} \int_{-L/2}^{L/2} dx\, (\delta \bar{V} - \delta\bar\mu)\left(u_n + v_n\right)\,,\quad M_{ln}=-\frac{i}{2} \mathcal{A}\int_{-L/2}^{L/2} dx\, (\delta \bar{V} - \delta\bar\mu) u_l v_n\,,\\
	A_{ln}&=& -\frac{i}{2} \mathcal{A}\int_{-L/2}^{L/2} dx\, (\delta \bar{V} - \delta\bar\mu) u_l u_n\quad\rm{and}\quad 
	B_{ln}= -\frac{i}{2} \mathcal{A}\int_{-L/2}^{L/2} dx\, (\delta \bar{V} - \delta\bar\mu) v_l v_n\,,
\end{eqnarray}
which for $\delta \bar{V} = m(\Phi_{0,\Omega} + a_{\Omega}x + \mathfrak{G} x^2/2)$ become
\begin{eqnarray}\label{eq:moments2}
	M_{0n}&\approx & im L^{3/2}\sqrt{\frac{N_a \zeta }{(\sqrt{2}n\pi)^3}}
	   \left(\left(1-(-1)^n\right)\frac{a_{\Omega}}{L}  - \left(1+(-1)^n\right) \frac{\mathfrak{G}_{\Omega}}{2}\right) \\
	M_{ln}&\approx & - A_{ln} \approx - B_{ln} \approx - im \frac{L^3(l^2+n^2)}{2\sqrt{2nl}(l^2-n^2)^2\pi^3\zeta}
	   \left(\left(1-(-1)^{l+n}\right)\frac{a_{\Omega}}{L}  - \left(1+(-1)^{l+n}\right) \frac{\mathfrak{G}_{\Omega}}{2}\right)\quad\rm{for}\quad l\neq n\,,\\
	M_{nn}&\approx & - A_{nn} \approx - B_{nn} \approx im\frac{\mathfrak{G}_{\Omega} L^3}{16\sqrt{2} n^3\pi^3\zeta}\,,
\end{eqnarray}
since $k_n\zeta \ll 1$. It is interesting to note that the absolute values of all transition amplitudes are monotonously decreasing with increasing $n$ and $l$ for fixed $n-l$. The maximum of $|M_{0n}|$ is reached for $n=1$ if $a_\Omega \neq 0$ and for $n=2$ if $a_\Omega = 0$. The maximum of $|M_{ln}|=|A_{ln}|=|B_{ln}|$ is reached for $n=1$ and $l=2$ if $a_\Omega \neq 0$ and for $n=1$ and $l=3$ if $a_\Omega = 0$. The maximum of $|M_{nn}|=|A_{nn}|=|B_{nn}|$ is reached for $n=1$. 

In the following, we will only consider processes on resonance; this means that $n_\Omega := L\Omega/(\pi c_0)$ is an integer. Then, we find for the time evolution
\begin{eqnarray}\label{eq:HIdeltaVmainres}
	  \nonumber \hat U^{\delta V}_\rm{int,res} &=&  \exp\Bigg[ (-1)^{n_\Omega}\alpha_{n_\Omega}(t)\left(\hat b^\dagger_{n_\Omega} - \hat b_{n_\Omega}  \right)  - \frac{1+(-1)^{n_\Omega}}{2}\frac{r_{n_\Omega/2}(t)}{2}\left( \hat b_{n_\Omega/2}^{\dagger 2} - \hat b_{n_\Omega/2}^2 \right)  \\
	 && -  \sum_{l<n_\Omega/2}(-1)^{n_\Omega} r_{l,n_\Omega-l}(t)\left(\hat b_l^\dagger \hat b_{n_\Omega-l}^\dagger  - \hat b_l \hat b_{n_\Omega-l} \right) +  \sum_{l>n_\Omega} (-1)^{n_\Omega} \Theta_{l,l-n_\Omega} \left(\hat b_l^\dagger \hat b_{l-n_\Omega}- \hat b_{l-n_\Omega}^\dagger \hat b_l  \right) \Bigg] \,,
\end{eqnarray}
where
\begin{equation}
	\alpha_{n}(t)=|M_{0n}|t/\hbar\,,\quad r_{n}(t)=2|M_{nn}|t/\hbar\,,\quad r_{l,n}(t)=2|M_{ln}|t/\hbar\quad\rm{and}\quad\Theta_{l,n}(t)=2|A_{ln}|t/\hbar\,.
\end{equation}
Note that the second term in Eq. (\ref{eq:HIdeltaVmainres}) only exists if $n_\Omega$ is even. In the following, we will discuss all the different terms in Eq. (\ref{eq:HIdeltaVmainres}), explain their meaning, investigate their effect on the phonon number and compare the results to those of Sec. \ref{sec:descriptionfluid}.

\subsection{Coherent displacement}

The first term in the exponent of the time evolution operator in Eq. (\ref{eq:HIdeltaVmainres}) would create a coherent state when the other terms are neglected. This coherent state leads to non-vanishing expectation values of the quadratures $(\hat\vartheta + \hat\vartheta^\dagger)/2$ and $-i(\hat\vartheta - \hat\vartheta^\dagger)/2$, which can be identified with $\sqrt{\rho_0}\alpha$ and $\sqrt{\rho_0}\phi$, respectively.  Therefore, the first term in Eq. (\ref{eq:HIdeltaVmainres}) can be identified with the direct driving $D_n$. Taking only  this term into account, the average number of phonons is given as $|\alpha_n|^2=|M_{0n}t/\hbar|^2$ and we recover the result in Eq. (\ref{eq:ncr}).

\subsection{Single-mode squeezing}

The second term in the exponent of the time evolution operator in Eq. (\ref{eq:HIdeltaVmainres}) corresponds to single-mode squeezing. When the system is parametrically driven on resonance $\Omega=2\omega_n$ and the other processes are neglected, we obtain for the time evolution of the ladder operators under the influence of single-mode squeezing $\hat b_n(t)=S_n^\dagger(r_n(t))\hat b_n S_n(r_n(t))$, where the squeeze operator
$S_n(r_n(t))$ is defined as
\begin{equation}	
	S_n(\xi)=\exp(-(\xi\hat b_n^{\dagger 2} - \xi^*\hat b_n^2)/2)
\end{equation} 
(see Sec. 16.1  and Sec. 2.7 of \cite{Scully:1997qua}). If the system starts in the vacuum, a squeezed state is created with squeezing parameter $r_n(t)$ and a number of squeezed phonons $N_{n,s} = \sinh^2 r_n$. This is the dynamical Casimir effect in Bose-Einstein condensates \cite{Jaskula:2012acou}. If we assume a general initial displaced squeezed state $|\alpha_n,\xi_{n,0}\rangle=D_n(\alpha_n)S_n(\xi_{n,0})|0\rangle$, where $D_n(\alpha_n)=\exp(\alpha_n \hat b^\dagger_n - \alpha_n^* \hat b_n)$ is the displacement operator, we find for the average number of phonons
\begin{equation}
		\bar{N}_{n}=|\cosh(r_{n,0})\sinh(r_n(t)) + e^{i\theta_{n,0}}\sinh(r_{n,0})\cosh(r_n(t))|^2 + |\alpha^*\cosh(r_n(t))-\alpha\sinh(r_n(t))|^2
\end{equation}
where $\xi_{n,0}=r_{n,0}e^{i\theta_{n,0}}$. For the particular case where the system starts in a coherent state of mode $n$, $D(\alpha_n)|0\rangle$ with parameter $\alpha_n$, a displaced squeezed state  $D(\beta_n)S_n(t)|0\rangle$ is created, where $\beta_n= \alpha_n \cosh r_n - \alpha_n^* \sinh r_n$. The total number of phonons in the displaced squeezed state is 
\begin{equation}
	\bar{N}_{n}=\sinh^2(r_n(t))+|\beta_n|^2\quad\rm{where}\quad |\beta_n|^2 =|\alpha^*\cosh(r_n(t))-\alpha\sinh(r_n(t))|^2\,.
\end{equation}
However, only $|\beta_n|^2$ phonons contribute to the mean field, i.e. to the quadratures $\alpha$ and $\phi$. For $\alpha_n^* = -\alpha_n$, we find that $|\beta_n|^2 =|\alpha_n|^2e^{2r_n}$ and we recover Eq. (\ref{eq:solpara}) for parametric driving. For $\alpha_n^*=\alpha_n$, we obtain $|\beta_n|^2 =|\alpha_n|^2e^{-2r_n}$, and we find that the squeeze operation pumps phonons out of the mean field. If the system starts from a squeezed vacuum state with squeezing parameter $r_{n,0}$ and $\theta_{n,0}=0$, a new squeezed state is created with an average number of squeezed phonons $\bar N_{n,\rm{s}} = \sinh^2 (r_{n,0} + r_n)$. 

Note that for $r_n\ll 1$ and an initial coherent state, we find 
\begin{equation}
	\delta \bar N_{n,\rm{c}}= |\beta_n|^2-|\alpha_n|^2 \approx \pm|\alpha_n|^2 2r_n = \pm \bar N_{n,\rm{c},0} 2r_n
\end{equation} for $\alpha_n^*=\mp \alpha_n$. Similarly, for the case of an initial squeezed state and for $r_n\ll 1$ and a number of initial phonons $\bar N_{n,\rm{s},0}=\sinh^2 (r_{n,0}) \gg 1$, we find that $\delta \bar N_{n,\rm{s}} = \bar N_{n,\rm{s}} - \bar N_{n,\rm{s},0} \approx  \bar N_{n,\rm{s},0}2r_n$. Hence, for a highly excited initial state, a small amplitude $M_{nn}$ still can lead to a measurable change in the average number of phonons. Such initial phonons may be created by applying an external electromagnetic linear or harmonic potential to the BEC superimposed with the trap potential by the same mechanisms that we consider here for the measurement process.

\subsection{Multi-mode squeezing}

Additionally to one mode squeezing, we find multi-mode squeezing with the third term in the exponent of the time evolution operator in Eq. (\ref{eq:HIdeltaVmainres}). In particular, this process can be induced by acceleration and the gravity gradient, while acceleration only creates phonon pairs in modes with different parity, the gravity gradient creates phonons in modes with the same parity. For small mode numbers $n$ and $l$ and $n+l$ even, the amplitude $M_{nl}$ will be of the same order as $M_{nn}$. However, the quotient of the amplitude $M_{nl}/M_{nn}$ increases with increasing mode numbers $l$ and $n$ when $n-l$ is kept constant. Therefore, to generate phonons with higher frequencies, multi-mode squeezing will be more efficient than one mode squeezing. In all cases, the resonance condition $\Omega\approx \omega_n + \omega_l $ has to be fulfilled to create phonons. Only taking multi-mode squeezing into account, the evolution of the creation and annihilation operators can be represented as $\hat b_j(t)=S_T(t)^\dagger \hat b_j S_T(t)$ using the two-mode squeeze operator 
\begin{equation}
	S_T(t)=\exp\left(-\sum_{l<n_\Omega/2} (-1)^{n_\Omega} r_{l,n_\Omega-l}(t)( \hat b_l^\dagger \hat b_{n_\Omega-l}^\dagger -  \hat b_l \hat b_{n_\Omega-l})\right)\,.
\end{equation}
Let us assume that the system starts in an initial displaced squeezed state $|\alpha_n,\xi_{n,0}\rangle$ with $n<n_\Omega$. Only taking multi-mode squeezing into account, we obtain the state $S_T(t)|\alpha_n,\xi_{n,0}\rangle$. We have 
\begin{eqnarray}
	S_T(t)^\dagger \hat b_n S_T(t)&=&\cosh(r_{n,n_\Omega-n}(t)) \hat b_n + (-1)^{n_\Omega}\sinh(r_{n,n_\Omega-n}(t))\hat b^\dagger_{n_\Omega-n}\\
	S_T(t)^\dagger \hat b_{n_\Omega-n} S_T(t)&=&\cosh(r_{n,n_\Omega-n}(t)) \hat b_{n_\Omega-n} + (-1)^{n_\Omega}\sinh(r_{n,n_\Omega-n}(t))\hat b^\dagger_n\,
\end{eqnarray}
and the total average number of phonons in mode $n$ and $n_\Omega-n$ becomes
\begin{equation}
	\bar N_{n}=\bar N_{n,0} + (\bar N_{n,0} + 1)\sinh^2(r_{n,n_\Omega-n}) \,\quad\rm{and}\quad \bar N_{n_\Omega-n}=(\bar N_{n,0} + 1)\sinh^2(r_{n,n_\Omega-n})\,,
\end{equation}
respectively, where $\bar N_{n,0} = \bar N_{n,\rm{s},0} + \bar N_{n,\rm{c},0}$ is the total number of phonons in the initial state, $\bar N_{n,\rm{s},0}=\sinh^2(r_{n,0})$ and $\bar N_{n,\rm{c},0}=|\alpha_n|^2$. Calculating the expectation value of the quadrature $\langle \alpha_n,\xi_{n,0}|S_T(t)^\dagger \frac{i}{2}(b_{n_\Omega-n}^\dagger - b_{n_\Omega-n})S_T(t)|\alpha_n,\xi_{n,0}\rangle$, we obtain the average number of phonons that contribute to the mean field in mode $n_\Omega-n$ as $\bar N_{n_\Omega-n,\rm{c}}=\bar N_{n,\rm{c},0}\sinh^2(r_{n,n_\Omega-n})$. For $r_{n,n_\Omega-n}< 1$, we find $\bar N_{n_\Omega-n,\rm{c}}\approx \bar N_{n,\rm{c},0}r_{n,n_\Omega-n}^2$. With the explicit expression for $r_{n,n_\Omega-n}$, we can compare this result to the driving of mode $n$ through the mode coupling discussed in Sec. \ref{sec:descriptionfluid}. From Eq. (\ref{eq:couplingcreated}), where we replace $n$ by $n_\Omega-n$ and $l$ by $n$, we recover $\bar N_{n_\Omega-n,\rm{c}} =  \bar N_{n,\rm{c},0}r_{n,n_\Omega-n}^2 $.

\subsection{Mode mixing}

The last term in the exponent of the time evolution operator in Eq. (\ref{eq:HIdeltaVmainres}) leads to mode mixing, which partially corresponds to the mode coupling process; the second term on the right hand side of Eq. (\ref{eq:harmosc}). The amplitude is the same as for the two-mode squeezing and the process will be efficient for high mode numbers $n$ and $l$ when $n-l$ is small. When only the mode mixing is taken into account and only resonant processes are considered, the time evolution can be described by a beam splitting operation $b_j(t)=B(t)^\dagger b_j B(t)$, where
\begin{equation}
	B(t)=\exp\left(\sum_{l>n_\Omega}(-1)^{2l-n_\Omega} \Theta_{l,l-n_\Omega}(t)  \left(b_l^\dagger b_{l-n_\Omega} - b_{l-n_\Omega}^\dagger b_l\right)\right)\,.
\end{equation}
This operator couples all modes with a distance $n_\Omega$. This means, that we obtain $n_\Omega-1$ systems of coupled modes \footnote{These systems of coupled modes can be seen as quantum Markov chains that are infinite on one side. $B(t)$ gives rise to the time evolution of an initial state on this Markov chain. The beam splitter operation is a completely positive map \cite{Aiello:2007lin} and Markov chains can be understood as completely positive maps between sites of a graph \cite{Gudder:2008qua}. Here we have a one-dimensional undirected graph with one end.}. 
Starting with a displaced squeezed initial state $|\alpha_n,\xi_{n,0}\rangle$ for a single-mode $n$ and assuming that $\Theta_{n,n\pm n_\Omega}(t)\ll 1$, we find that 
\begin{equation}
	\bar N_{n}=\langle \alpha_n,\xi_{n,0}|b_n^\dagger(t) b_n(t)  |\alpha_n,\xi_{n,0}\rangle \approx \bar N_{n,0} (1-2(\Theta_{n,n- n_\Omega}^2 + \Theta_{n,n+ n_\Omega}^2))\,,
\end{equation}
where $\bar N_{n,0} = \bar N_{n,\rm{s},0} + \bar N_{n,\rm{c},0}$ is the total number of phonons in the initial state.

\subsection{Quantum Cram\'er-Rao Bound}

From our analysis above we see that higher creation rates of phonons are obtained for initially excited states. From the created phonons the gravitational field amplitudes $a_\Omega$ and $\mathfrak{G}_\Omega$ can be inferred. Therefore, optimal precision for the measurement of the gravitational field via the creation of phonons is obtained for excited states, which we can call probe states. The estimation via the squeezing effect represented by $M_{nn}$ and $M_{ln}$ is sometimes called a squeezing channel. The estimation via the mode mixing represented by $A_{ln}$ and $B_{ln}$ is called a mode mixing channel. It was shown in \cite{Safranek:2016gyk} that the optimal estimation of a parameter via either of the two channels is obtained for a probe state where all particles are squeezed particles. Here, optimal means highest precision per phonon in the probe state. In the following, we will give this optimal precision in terms of the Quantum Cram\'er-Rao Bound, which gives an upper limit for the precision of the estimation of a parameter via an estimation channel for a specific initial state for all possible measurements. We obtain the limit for the absolute precision for a measurement of the driving parameter $\epsilon\in \{a_\Omega,\mathfrak{G}_\Omega\} $ as
\begin{equation}\label{eq:QCRB}
	 \sqrt{\Delta^\rm{CR}_\epsilon} \ge \frac{1}{\sqrt{\#_\rm{rep} I_\epsilon}} \,,
\end{equation}
where $\#_\rm{rep}$ is the number of consecutive independent measurements and $I_\epsilon$ is the Quantum Fisher Information for the driving parameter $\epsilon$.

Let us assume that we start with two non-entangled modes each in a squeezed probe state with squeezing parameters $r_{0,1}$ and $r_{0,2}$. The Quantum Fisher Information for the optimal estimation of the driving parameter $\epsilon$ via the two-mode squeezing channel is given as $I_\epsilon=(2(2N_{s,0}+1)\delta r(t)/\epsilon)^2$ for $r_{0,1}=r_{0,2}$, where $\delta r(t)=2|M_{n_1n_2}|t/\hbar$ and $N_{s,0}=\sinh^2r_{0,1}+\sinh^2r_{0,2}$ is the initial number of squeezed phonons \cite{Safranek:2016gyk}. The Quantum Fisher Information for the optimal estimation via the mode mixing channel is $I_\epsilon=4N_{s,0}(N_{s,0}+2)(2|A_{n_1 n_2}|t/(\hbar\epsilon))^2$ \cite{Safranek:2016gyk}. Both expressions correspond to the Heisenberg limit and represent an upper bound for the achievable sensitivity if the probe state is set up as described above.

In the next section, we will investigate the sensitivity for the estimation of acceleration and gravity gradient for examples of experimental parameters.

%%%%%%%%%%%%%%%%%%%%%%%%%%%%%%%%%%%%%%%%%%%%%%%%%

\section{Experimental parameters and measurement sensitivity}
\label{sec:exppar}

In this section, we give necessary experimental parameters for the measurement of the gravitational field of an oscillating mass via phonon creation in a BEC. To provide an example, let us consider rubidium BECs and ytterbium BECs. For Rb-87, we have an interaction constant of $\lambda_\rm{Rb}\approx 1.3\times 10^{-7}\rm{m}$ (calculated from the measured scattering length $a_\rm{scatt}=\lambda/8\pi\approx 98 a_0$ reported in \cite{Egorov:2013meas}, where $a_0$ is the Bohr radius). The mass of a Rb-87 atom is $1.44\times 10^{-25}\,\rm{kg}$. The scattering length of Yb-168 can be found in \cite{Kitagawa:2008two} and leads to a interaction constant of approximately $\lambda_\rm{Yb}=3.35\times 10^{-7}$. 
The mass of a Yb-168 atom is $2.79\times 10^{-25}\,\rm{kg}$. In the following, we will assume that the cross section of the BEC in the $y$-$z$-plane is circular. 

The experimental time $t_\rm{exp}$ is limited by the half-life of the BEC and the half-life of the phonons. We assume that the half-life of the phonons, which is proportional to the inverse damping rate, is much larger than the half-life of the BEC. We find that this assumption is met for the experimental parameters chosen. Therefore, the half-life is the only limit for $t_\rm{exp}$ in the following consideration. In Section 5.4 of \cite{Pethick:2002bose} and in \cite{Norrie:2006thr,Moerdijk:1996dec} it is shown that $d\rho(t)/dt = -D\rho(t)^3$, where $D$ is the decay constant. This implies a quadratic dependence of the half-life on the inverse density. More precisely $t_\rm{hl}=3/(2D\rho^2)$. For example, in an experiment with rubidium atoms \cite{Soding:1999three}, the corresponding decay constant was found to be $1.8\times 10^{-29}\,\rm{cm}^6\rm{s}^{-1}$. In an experiment \cite{Takuso:2003spin} with ytterbium an even smaller decay constant of $4\times 10^{-30}\,\rm{cm}^6\rm{s}^{-1}$ was found.
Therefore, we assume that the density is less or equal to $10^{13}\,\rm{cm}^{-3}$ in order to achieve a half-life of at least $t_\rm{hl}= 100\,\rm{s}$. This allows the assumption that the number of atoms can be considered to be constant during a single run of the experiment for an interaction time of $t_\rm{exp}= 10\,\rm{s}$. 
As we discussed in Sec. \ref{sec:damping}, the decay of the BEC leads to a damping of the phonons inside the BEC, with a value for the damping rate $\gamma^\rm{loss}$ that approximately matches the value of the BEC decay rate. From the discussion above, we find that $\gamma^\rm{loss}\approx 10^{-2}\,\rm{s}^{-1}$ is a conservative assumption. For all the experimental parameters that we will consider in the following, we find that the Landau damping rate $\gamma_n^\rm{L}$ is of the order of $10^{-3}\,\rm{s}^{-1}$ and Beliaev damping is significantly smaller. Therefore, the total phonon damping rate is dominated by the atom loss, we can set $\gamma_n=\gamma^\rm{loss}$, and we obtain $\gamma_n^{-1}\gg t_\rm{exp}$. We assume that $\Omega \ge 2\pi\times 5/t_\rm{exp}= 2\pi \times 0.5\,\rm{Hz}$, which means that at least about $5$ cycles of the driving oscillation elapse during one run of the experiment. To fulfill the condition $k_B T \ll \mu$, we consider a temperature of $1\,\rm{nK}$. In a uniform BEC this leads to a relative depletion\footnote{For the depletion, we used the formula (4.50) in \cite{Pitaevskii:2003bose}: $(\rho_0(T)-\rho_0(T=0))/\rho_0(T=0)=-m(k_B T)^2/(12\rho c_0\hbar^3)$, where $\rho_0(T)$ and $\rho_0(T=0)$ are the density of atoms in the ground state at temperature $T$ and $T=0$, respectively.} of the density of atoms in the ground state of the order of $10^{-4}$. Accordingly, less than $10^{-3}N_a$ atoms are in thermally excited states. The back action of these atoms on the condensate leads to a stationary deformation of the order parameter and the number of phonons created due to the oscillating gravitational field is changed slightly. Hence, thermal depletion is a small effect for the parameters considered in this article that can be neglected.

\subsection{Phonon creation due to direct driving}

In the following, we want to find experimental parameters that allow for the detection of gravitational acceleration by measurement of phonons created via direct driving. To achieve the maximal creation rate, the mode number $n$ has to be as small as possible. Therefore, we consider $n=1$ for the creation of phonons by direct driving due to the oscillating acceleration.
From the expression for the frequency $\omega_n=c_0 k_n$ and the definition of the speed of sound in Eq. (\ref{eq:soundspeed}), we find
\begin{equation}\label{eq:Lfix}
	L=\frac{\pi }{\omega_1}\frac{\hbar}{m}\sqrt{\frac{\lambda \rho_0}{2}}\,.
\end{equation}
The resonance condition is $\omega_1=\Omega$ and we assumed that $\Omega \ge 2\pi \times 0.5\,\rm{Hz}$. This condition is fulfilled if we choose $L=200\,\rm{\mu m}$, which corresponds to $\omega_1= 2\pi \times 1.5\,\rm{Hz}$ for rubidium and $\omega_1= 2\pi \times 1.2\,\rm{Hz}$ for ytterbium.

For a successful detection, we need the signal to noise ratio $\mathcal{R}_\rm{SNR}=a_{\omega_n}/\Delta a_{\omega_n}^\rm{tot}$ to be much larger than one, where $\Delta a_{\omega_n}^\rm{tot}$ is the variance of the fluctuations of the measurement signal. Let us assume that the length $L$ and the transversal cross section $\mathcal{A}$ of the BEC can be specified with high precision. We find three main sources of fluctuations that contribute to $\Delta a_{\omega_n}^\rm{tot}$: the precision of detection $\Delta N_\rm{n,\rm{det}}$ for the number of phonons in a mode, the fluctuations of the number of thermal phonons in the BEC $\Delta N_\rm{n,\rm{th}}$ and the fluctuations of the number of atoms in the BEC $\Delta N_a$. The number of created phonons can be inferred from the number of detected phonons by subtracting the number of thermal phonons; $N_{n,cr}=N_\rm{n,det} - N_\rm{n,th}$. Inverting Eq. (\ref{eq:ncr}), we find by Gaussian error propagation
\begin{equation}\label{eq:signaltonoise}
	\mathcal{R}_\rm{SNR,n} \approx \sqrt{\#_\rm{rep}}\left(\frac{1}{4\bar{N}_{n,cr}^2}\left(\Delta N_\rm{n,th}^2 + \Delta N_\rm{n,det}^2\right) + \frac{1}{16\bar N_a^2} \Delta N_a^2\right)^{-1/2}
\end{equation}
where $\#_\rm{rep}$ is the number of repetitions of the experiment and $\bar{N}_{n,cr}=\bar{N}_{n,\rm{c},t\ll \gamma_n^{-1}}$ is the average number of created phonons in mode $n$. Note that $N_a$ enters the right hand side of Eq. (\ref{eq:ncr}) directly and through $\zeta$ which leads to the factor $1/16$ in front of $\Delta N_a^2$ in Eq. (\ref{eq:signaltonoise}). In state of the art experiments \cite{Leanhardt:2003coo}, the temperature of the BEC varies by about $20\,\%$ between two runs. We assume that the temperature is about $1\,\rm{nK}$, which implies that $k_B T\gg \hbar \omega_n$. Hence, the Bose-Einstein statistics tells us that the value for the variance of the number of thermal phonons is approximately equivalent to the value for the average number of phonons. We find $\Delta N_\rm{n,th}/\bar{N}_\rm{n,th} \sim 1$, where $\bar{N}_\rm{n,th}=k_B T/(\hbar\omega_n)$ is the average number of thermal phonons in mode $n$. For the detection error we assume single phonon sensitivity independently of the mode number, i.e. $\Delta N_\rm{n,det}=1$. We will discuss possibilities of a detection process in Sec. \ref{sec:conclusions}. Furthermore, we assume that the number of atoms varies by about $10\%$ between two experiments, which means $\Delta N_a/\bar{N}_a\sim 0.1$. Since each experiment takes about $10\,\rm{s}$, we can consider a number of repetitions of $\#_\rm{rep}=10^4$, which corresponds to two days of consecutive measurements. For example for $\omega_1 \sim 2\pi \times 1\,\rm{Hz}$, we obtain that $\bar{N}_\rm{n,th} \sim 20$. Then, the thermal fluctuations are the main source of fluctuations, and we find that the creation of a single phonon by the gravitational field would be sufficient to reach a signal to noise ratio of the order of $10$. In the following, we assume $\mathcal{R}_\rm{SNR,1}=10$, and we give the experimental parameters necessary to achieve this goal.
\begin{figure}
	\begin{tabular}{ |c|c|c|c|c|c||c|c|c|c|c|c|c|c|  }
	 \hline
	 atom species & $M$ & $R_\rm{min}$ & $\delta_R$ & $a_{\Omega}$ & $\Omega/2\pi$ & $N_a$  & $L/\zeta$ & $d/L$ & $\bar{N}_{1,cr}$ & $N^{\rm{lim},1}_{2}$  & $\bar{N}_\rm{1,th}$  \\
	 \hline
	 Rb-87   & 200 g & 1 mm & 2 mm & 2$\times10^{-8}\,\rm{ms^{-2}}$ & $1.5\,\rm{Hz}$ & 9$\times10^5$ & 230 & 0.12 & 0.7 & 1.3 & 14 \\
	 Yb-168  & 200 g & 1 mm & 2 mm & 2$\times10^{-8}\,\rm{ms^{-2}}$ & $1.2\,\rm{Hz}$ & 5$\times10^5$ &  370 & 0.08 &  0.9 & 0.16 & 17 \\ 
	 Rb-87   & 0.2 g & 0.1 mm & 0.2 mm & 2$\times10^{-9}\,\rm{ms^{-2}}$ & $1.5\,\rm{Hz}$ & 1$\times10^8$ & 230 & 1.4 &  0.7 & 180 & 14 \\
	 Yb-168  & 0.2 g & 0.1 mm & 0.2 mm & 2$\times10^{-9}\,\rm{ms^{-2}}$ & $1.2\,\rm{Hz}$ &  6$\times10^7$ & 370 & 1 &  0.9 & 23 & 17 \\ 
	 \hline
	\end{tabular}
	\caption{\label{fig:table1} COHERENT/DIRECT DRIVING: This table shows some generic values for the experimental parameters necessary to detect phonons in a BEC created by direct driving due to the oscillating gravitational acceleration of amplitude $a_{\Omega}$ induced by a small oscillating sphere of mass $M$ with a signal to noise ratio of the order $10$. It is assumed that the density of the BEC is $\rho=10^{13}\,\rm{cm^{-3}}$, the length of the uniform trap potential well is $L=200\,\rm{\mu m}$, the temperature of the BEC is $T=1\,\rm{nK}$ and the measurement precision is of the order of a single phonon. The interaction time for each experiment is assumed to be $t_\rm{exp}=10\,\rm{s}$. In each case, about $10^4$ repetitions of the experiment are considered. Under these conditions, the minimal distance $R_\rm{min}$ to the BEC, the oscillation amplitude $\delta_R$, the frequency of the driving $\Omega/2\pi$, the number of atoms $N_a$ in the BEC, the ratio of length of the BEC and healing length $L/\zeta$ and the ratio of the length and the diameter $d/L$ of the trap potential are given. Additionally, the table shows the average number of created phonons, the value for the number of phonons at which the inter-mode coupling becomes significant $N^{\rm{lim},1}_{2}$ and the average number of thermal phonons. }
\end{figure}
The necessary number of created phonons for the detection process $\bar{N}_{n,cr}$ can be used to give a lower bound for the number of atoms in the BEC by using Eq. (\ref{eq:Lfix}) and Eq. (\ref{eq:ncr}), which leads to
\begin{equation}\label{eq:ncrinv}
	N_a = \frac{(n\pi)^2 \hbar\omega_n \bar{N}_{n,cr} }{m t^2 a_{\Omega}^2} \,.
\end{equation}
Eq. (\ref{eq:ncrinv}) gives a lower bound because $\omega_n$ is bounded from below and all other parameters in Eq. (\ref{eq:ncrinv}) can be fixed a priori: $\lambda$ and $m$ are fixed by choosing an atom species and the time is the experimental time $t_\rm{exp}$.

The fixed length and the fixed number of atoms can be used to fix the ratio between the transversal diameter $d$ and the length of the BEC. We obtain
\begin{equation}\label{eq:quotient}
	\frac{d}{L}=2\sqrt{\frac{N_a}{\pi\rho_0 L^3}}\,.
\end{equation}
The density should be as large as possible in order to keep $d$ small. Therefore, we match the upper bound for the density given above and set $\rho = 10^{13}\,\rm{cm}^{-3}$. 

In Fig. \ref{fig:table1}, a table can be found in which some example values for experimental parameters are listed that could be used for the detection of the gravitational field of an oscillating gold or tungsten sphere of mass $M$ for the two cases of $M=200\,\rm{g}$ and $M=0.2\,\rm{g}$. Values for the parameters are given for both a rubidium and an ytterbium BEC. For a source mass of $200\,\rm{g}$, we find that the phonon creation should be observable with state of the art technology. In the case of ytterbium, the number of phonons for which the mode coupling becomes significant is smaller than the number of created phonons. This suggests that a regime can be reached in which mode coupling and parametric driving may supply an alternative detection scheme. We will investigate this possibility in the next subsection. If we repeat the above calculations for the creation of phonons due to the oscillating gravity gradient, we find that we would need about $10^{10}$ atoms for the detection using the direct driving process. This is experimentally out of reach at the moment. Mode coupling and parametric driving will be much more efficient for gradiometry.

\subsection{Phonon creation due to squeezing}

In Sec. (\ref{sec:coherent}), we discussed the creation of coherent phonons in an initial coherent state and the creation of additional squeezed phonons in an initial squeezed state due to the squeezing processes. The first situation corresponds to the creation of phonons in the mean field due to the parametric driving of the BEC by the oscillating gravitational field discussed in Sec. \ref{sec:resdriv}. We found that the total average number of phonons that are created on resonance in the mode $n$ by multi-mode squeezing is given as $\bar N_{n,cr}=\delta\bar N_{n} = \bar N_{n,0} r_{n,n_\Omega-n}^2$ for $r_{n,n_\Omega-n}=2|M_{n,n_\Omega-n}|t/\hbar < 1$, where $\bar N_{n,0}$ was the initial number of phonons (squeezed plus coherent) in mode $n$ and $n_\Omega = \Omega L/(\pi c_0)$ is an integer since we consider resonant driving.

The amplitude $|M_{n,n_\Omega-n}|$ can be maximized by choosing $L$ large and $|n^2-(n_\Omega-n)^2|$, $n$, $n_\Omega$ and $\zeta$ as small as possible. In contrast to the direct driving, the number of atoms in the BEC does not appear in the number of created phonons due to squeezing. Since $\zeta=1/\sqrt{\lambda \rho_0}$, the healing length can be minimized by choosing for a density of $\rho_0 = 10^{13}\,\rm{cm}^{-3}$, which we identified as the maximum when an experimental time of at least $10\,\rm{s}$ is to be obtained. Note that only gravitational acceleration contributes for $n_\Omega$ odd and only the gravity gradient contributes for $n_\Omega$ even. We can minimize $|n^2-(n_\Omega-n)^2|$ by choosing $n=(n_\Omega+1)/2$ if $n_\Omega$ is odd and $n=n_\Omega/2+1$ if $n_\Omega$ is even.  We consider $n=2$ and $n_\Omega=3$ for the measurement of the acceleration and $n=3$ and $n_\Omega=4$ for the measurement of the gravity gradient in the following. For the detection of the oscillating acceleration, we fix the length of the BEC to $L=200\,\rm{\mu m}$. For the detection of the oscillating gravity gradient, we have to increase the length to obtain more realistic values for the other experimental parameters; we consider $L=500\,\rm{\mu m}$.

Now, we can calculate the number of initial phonons $\bar N_{n,0}$ that are necessary to create $\bar{N}_{n,cr}$ phonons on average in a single experiment, which we need to achieve a signal to noise ratio of the order of $10$ after $10^4$ repetitions of the experiment. We find
\begin{equation}
	\bar N_{n,0} = \frac{\bar{N}_{n,cr}}{r_{n-n_\Omega,n}^2}\,.
\end{equation}
The relation between the signal to noise ratio and $\bar{N}_{n,cr}$ can be derived from Eq. (\ref{eq:couplingcreated}) as
\begin{equation}\label{eq:signaltonoisecoupling}
	\mathcal{R}_\rm{SNR,n} \approx \sqrt{\#_\rm{rep}}\left(\frac{1}{4\bar{N}_{n,cr}^2}\left(\Delta N_\rm{n,th}^2 + \Delta N_\rm{n,det}^2\right) + \frac{1}{4\bar N_a^2} \Delta N_a^2 + \frac{1}{4\bar N_{l,\rm{c},0}^2} \Delta N_{l,\rm{c},0}^2\right)^{-1/2}
\end{equation}
which differs from Eq. (\ref{eq:signaltonoise}) by a different proportionality to $\Delta N_a$ and the additional contribution of the fluctuation of the number of initial coherent phonons $\Delta N_{l,\rm{c},0}$. Let us assume that $\Delta N_{l,\rm{c},0}$ is about one per cent of the total number of initial phonons $N_{l,\rm{c},0}$. Then, the thermal fluctuations are again the most significant source of uncertainty. 

Below Eq. (\ref{eq:nofph}), we identified the condition $\bar N_{n,\rm{c}}\le 10^{-2}n\pi\zeta N_a/(2\sqrt{2}L)$ in order to keep the phase perturbation $\phi\le 0.1$. The same condition applies to squeezed phonons since the variance of the perturbation must be a perturbation to justify the approximations we used to derive the results presented in this article. By setting $\bar N_{n,0}=\bar N_{n,\rm{c}}$, we obtain for the minimum number of atoms
\begin{equation}
	N^\rm{min}_a = 10^{2}\frac{2\sqrt{2} L \bar N_{n,0}}{n\pi\zeta}\,.
\end{equation}
The results for the above parameters are presented in the table in Fig. \ref{fig:table2}.
\begin{figure}
	\begin{tabular}{ |c|c|c|c|c|c||c|c|c|c|c|c|c|c|  }
	 \hline
	 atom species & $M$ & $R_\rm{min}$ & $\delta_R$ & $a_{\Omega}$ & $\Omega/2\pi$ & $N^\rm{min}_a$  &   $r_{12}$  &  $N_{2,\rm{cr}}$  &  $\bar N_{2,0}$ & $\bar N_{2,\rm{th}}$ \\
	 \hline
	 Rb-87   & 200 g & 1 mm & 2 mm & 2$\times10^{-8}\,\rm{ms^{-2}}$ & $4.4\,\rm{Hz}$ & 5 $\times10^4$  & 0.3 & 0.4 & 5  & 4 \\
	 Yb-168  & 200 g & 1 mm & 2 mm & 2$\times10^{-8}\,\rm{ms^{-2}}$ & $3.7\,\rm{Hz}$ & 1$\times10^4$   & 0.8  & 0.4 & 1 & 4 \\ 
	 Rb-87   & 0.2 g & 0.1 mm & 0.2 mm & 2$\times10^{-9}\,\rm{ms^{-2}}$ & $4.4\,\rm{Hz}$ & 7$\times10^6$  & 0.02  & 0.4 & 700 & 4\\
	 Yb-168  & 0.2 g & 0.1 mm & 0.2 mm & 2$\times10^{-9}\,\rm{ms^{-2}}$ & $3.7\,\rm{Hz}$ &  1$\times10^6$ & 0.07   & 0.4 & 100 & 4 \\ 
	 \hline
	\end{tabular}
	
\vspace{0.3cm}	
		\begin{tabular}{ |c|c|c|c|c|c||c|c|c|c|c|c|c|c|  }
	 \hline
	atom species & $M$ & $R_\rm{min}$ & $\delta_R$ & $\mathfrak{G}_{\Omega}$ & $\Omega/2\pi$ & $N^\rm{min}_a$  &   $r_{13}$  &  $N_{3,\rm{cr}}$  &  $\bar N_{3,0}$ & $\bar N_{3,\rm{th}}$  \\
	 \hline
	 Rb-87   & 200 g & 1 mm & 2 mm & 2$\times10^{-6}\,\rm{s^{-2}}$ & $2.4\,\rm{Hz}$ & 1$\times10^8$ &  0.008 & 0.6 &  $9\times 10^3$ & 4\\
	 Yb-168  & 200 g & 1 mm & 2 mm & 2$\times10^{-6}\,\rm{s^{-2}}$ & $2\,\rm{Hz}$ & 3$\times10^7$ &  0.02  & 0.7 & $1\times 10^3$  & 5 \\ 
	 Rb-87   & 0.2 g & 0.1 mm & 0.2 mm & 1.2$\times10^{-6}\,\rm{s^{-2}}$ & $2.4\,\rm{Hz}$ & 4$\times10^8$ & 0.005 &  0.6  & $2\times 10^4$ & 4 \\
	 Yb-168  & 0.2 g & 0.1 mm & 0.2 mm & 1.2$\times10^{-6}\,\rm{s^{-2}}$ & $2\,\rm{Hz}$ & 8$\times10^7$ & 0.01 & 0.7  & $3\times 10^3$ & 5 \\ 
	 \hline
	\end{tabular}
	\caption{\label{fig:table2} TWO-MODE SQUEEZING: These tables show some generic values for the experimental parameters necessary to detect the phonons in a BEC created by two-mode squeezing due to the oscillating gravitational acceleration of amplitude $a_{\Omega}$ and the oscillating gravity gradient of amplitude $\mathfrak{G}_{\Omega}$ induced by small oscillating sphere of mass $M$ with a signal to noise ratio of the order of $10$. It is assumed that the density of the BEC is $\rho=10^{13}\,\rm{cm^{-3}}$, the temperature of the BEC is $T=1\,\rm{nK}$ and the measurement precision is of the order of a single phonon. Furthermore, the length of the uniform trap potential well is $L=200\,\rm{\mu m}$ for the consideration of $a_{\Omega}$ and $L=500\,\rm{\mu m}$ for the consideration of $\mathfrak{G}_{\Omega}$. The  interaction time for each experiment is assumed to be $t_\rm{exp}=10\,\rm{s}$. In each case, about $10^4$ repetitions of the experiment are considered. Under these conditions, the minimal distance $R_\rm{min}$ to the BEC, the oscillation amplitude $\delta_R$, the frequency of the driving $\Omega/2\pi$ and the minimal number of atoms $N_a^\rm{min}$ in the BEC are given. Additionally, the tables show the values for the squeezing parameter $r_{1,2}$ and $r_{1,3}$, the average number of created phonons $N_{2,\rm{cr}}$ and $N_{3,\rm{cr}}$, the necessary number of initial phonons $\bar N_{2,0}$ and $\bar N_{3,0}$ and the average number of thermal phonons. }
\end{figure}

Additionally to multi-mode squeezing, we can also consider the utility of single-mode squeezing for the detection of the gravitational field of the oscillating mass. From Eq. (\ref{eq:moments2}), we see that single-mode squeezing can only be induced by an oscillating gravity gradient. The resulting parameters for single-mode squeezing can be found in Fig. \ref{fig:table3}. We see that single-mode squeezing gives better parameters than multi-mode squeezing for the measurement of the gravity gradient. This is because, for small $r_{n_{\Omega/2}}(t)$ and $r_{n,n_\Omega-n}(t)$, the number of created phonons is directly proportional to $r_{n_{\Omega/2}}(t)$ for single-mode squeezing, while it is proportional to the square of $r_{n,n_\Omega-n}(t)$ for multi-mode squeezing. For small wave numbers $n$ and $n_\Omega$, $r_{n_{\Omega/2}}(t)$ and $r_{n,n_\Omega-n}(t)$ are approximately of the same order.
\begin{figure}
\begin{tabular}{ |c|c|c|c|c|c||c|c|c|c|c|c|c|c|  }
	 \hline
	atom species & $M$ & $R_\rm{min}$ & $\delta_R$ & $\mathfrak{G}_{\Omega}$ & $\Omega/2\pi$ & $N^\rm{min}_a$  &   $r_{1}$  &  $N_{1,\rm{cr}}$  &  $\bar N_{1,0}$ & $\bar N_{1,\rm{th}}$  \\
	 \hline
	 Rb-87   & 200 g & 1 mm & 2 mm & 2$\times10^{-6}\,\rm{s^{-2}}$ & $1.2\,\rm{Hz}$ & 4$\times10^6$ &  0.01 & 1.8 &  $80$ & 35\\
	 Yb-168  & 200 g & 1 mm & 2 mm & 2$\times10^{-6}\,\rm{s^{-2}}$ & $1\,\rm{Hz}$ & 3$\times10^6$ &  0.03  & 2.1 & $30$  & 42 \\ 
	 Rb-87   & 0.2 g & 0.1 mm & 0.2 mm & 1.2$\times10^{-6}\,\rm{s^{-2}}$ & $1.2\,\rm{Hz}$ & 7$\times10^6$ & 0.007 &  1.8  & $130$ & 35 \\
	 Yb-168  & 0.2 g & 0.1 mm & 0.2 mm & 1.2$\times10^{-6}\,\rm{s^{-2}}$ & $1\,\rm{Hz}$ & 4$\times10^6$ & 0.02 & 2.1  & $50$ & 40\\
	 \hline
	\end{tabular}
	\caption{\label{fig:table3}   SINGLE-MODE SQUEEZING: This table shows some generic values for the experimental parameters necessary to detect phonons in a BEC created by single-mode squeezing due to the oscillating gravity gradient of amplitude $\mathfrak{G}_{\Omega}$ induced by a small oscillating sphere of mass $M$ with a signal to noise ratio of the order of $10$. It is assumed that the density of the BEC is $\rho=10^{13}\,\rm{cm^{-3}}$, the temperature of the BEC is $T=1\,\rm{nK}$ and the measurement precision is of the order of a single phonon. Furthermore, the length of the uniform trap potential well is $L=500\,\rm{\mu m}$. The interaction time for the experiments is assumed to be $t_\rm{exp}=10\,\rm{s}$ and about $10^4$ repetitions of the experiment are considered. Under these conditions, the minimal distance $R_\rm{min}$ to the BEC, the oscillation amplitude $\delta_R$, the frequency of the driving $\Omega/2\pi$ and the minimal number of atoms $N_a^\rm{min}$ in the BEC are given. Additionally, the tables show the values for the squeezing parameter $r_{1}$, the average number of created phonons $N_{1,\rm{cr}}$, the necessary number of initial phonons $\bar N_{1,0}$ and the average number of thermal phonons.}
\end{figure}

\subsection{Measurement via mode mixing}

The remaining channel that can be used for a measurement is mode mixing; phonons in one mode will be transferred to another mode due to the oscillating gravitational field. Since the mode frequencies are equidistant, if there is a driving frequency $\Omega = \omega_{n_2} - \omega_{n_1}$, there is a resonance for $\omega_{n_{2+l}}=\omega_{n_2} + \Omega l$ and $\omega_{n_{1+l}}=\omega_{n_1} + \Omega l$. Therefore, phonons that are transferred to a higher mode from an initially excited mode will not stay in that mode but will be transferred up the whole cascade of resonant modes. Hence, starting from an excited state in one mode, we could measure the decrease of the number of phonons in this mode or the increase of the total number of phonons in all modes. 

From the amplitude in Eq. (\ref{eq:moments2}), we see that we can apply the same arguments as for the multi-mode squeezing, the only difference being that we consider the two modes $n$ and $n-n_\Omega$. If we consider $n=3$ and $n_\Omega=1$ for measurement of the gravitational acceleration, we obtain less favorable experimental parameters than those in the table in Fig. \ref{fig:table2} above. If we consider $n=3$, $n_\Omega=2$ for the measurement of the gravity gradient, we recover the same parameters as in the table in Fig. \ref{fig:table2} with the exception of the driving frequency which is decreased.

\subsection{Quantum Cram\'er-Rao Bound}

We can obtain an upper bound for the sensitivity of the measurement of oscillating gravitational fields using phonons in BECs by considering the Quantum Cram\'er-Rao Bound that we introduced in Eq. (\ref{eq:QCRB}). Considering multi-mode squeezing, the experimental parameters given in Fig. \ref{fig:table2} and about $1000$ initial squeezed phonons, we obtain an absolute error bound of the order of $10^{-13}\,\rm{ms^{-2}}$ ($L=200\,\rm{\mu m}$) for the measurement of acceleration and $10^{-10}\,\rm{s}^{-2}$ ($L=500\,\rm{\mu m}$) for the measurement of the gravity gradient. Comparing with the table in Fig. \ref{fig:table2}, we find that we could, in principle, measure the gravitational field of a $200\,\rm{mg}$ mass with a relative precision of $10^{-4}$.

\section{Conclusions and Discussions}
\label{sec:conclusions}

The necessary experimental parameters for the measurement of the gravitational field of an oscillating sphere of mass $M=200\,\rm{g}$ due to the direct driving of phonon modes in a BEC with a signal to noise ratio of the order of $10$ seem to be ambitious but not inaccessible (see the first table in Fig. \ref{fig:table1} for details). 
State of the art experiments with ultracold rubidium BECs (at about $1\,\rm{nK}$) use a number of atoms of the order of $10^5$ \cite{Gaunt:2013bose,Muentinga:2013int,Kovachy:2015mat,Hartwig:2015test} and atom numbers of the order of $10^6$ are planned for a new generation of experiments \cite{Schuldt:2014ofa,Aguilera:2014ste,Williams:2016qua}. In Sec. \ref{sec:exppar}, we argued that the interaction time for a single experiment of the order of $10\,\rm{s}$ can be achieved by choosing a low atom density of the order of $10^{13}\,\rm{cm}^{-3}$. The parameters for the case of $M=0.2\,\rm{mg}$ are out of range of state of the art experiments; the number of $10^8$ atoms necessary to achieve detection, with a signal to noise ratio of 10, is not obtainable. Nevertheless, this parameters may be achievable in the future.

Besides phonon creation due to direct driving, we investigated phonon creation due to parametric driving resulting in squeezing and mode mixing. This driving mechanism turns out to be of advantage when the phonon modes are initially in an excited state. Such initial excitations may be created by adding an oscillating external electromagnetic linear or harmonic potential to the already existing BEC trap potential. Then, phonons would be created by the mechanisms that we consider here for the measurement process. For example, this could be the direct driving or parametric driving from the vacuum, where the latter is equivalent to the dynamical Casimir effect in Bose-Einstein condensates \cite{Jaskula:2012acou}. See also \cite{Robertson:2018gwi}, where parametric amplification of excitations of phonons modes due to a modulation of the transverse trapping frequency of a BEC is discussed in detail.

We gave necessary experimental parameters for the measurement of the gravitational field of an oscillating massive sphere using parametric driving in Fig. \ref{fig:table2}. Firstly, it is interesting to note that our theoretical considerations predict that ytterbium would perform much better than rubidium.  Secondly, even for a few initial phonons, the parametric driving mechanism is more efficient than the direct driving as a lot less atoms are needed in the BEC to achieve a signal to noise ratio of the order of $10$. This is particularly useful for the measurement of the gravitational field of smaller masses; the acceleration due to a sphere of $200\,\rm{mg}$ can be measured with $10^6$ atoms when the initial state contains about $100$ coherent phonons. The measurement of the gravity gradient of small spheres with masses of the order of $200\,\rm{g}$ or less using direct driving is completely out of reach. However, it can be achieved with parametric driving with a BEC consisting of about $10^6$ atoms and $10$ to $100$ initial phonons. In this article, we assumed that the BEC is always much smaller than the distance between its center and the center of the source mass. It would be interesting to relax this condition in a future investigation. On the one hand, decreasing the distance between the source and the BEC beyond that limit may lead to further improvement in the measurement sensitivity. On the other hand, measurements close to the surface of small masses may allow for experiments to search for hypothetical fifth forces or to measure Casimir-Polder forces.
Finally, we calculated the Quantum Cram\'er-Rao Bound for the measurement precision considering the same experimental parameters. Hence, there seems to be a lot of potential for improvement by taking the insights of quantum metrology into account. This potential may become accessible through measurement schemes other than just the direct counting of created phonons such as, for example, a homodyne measurement of phonon modes.

In the this article, we assumed that the measurement technique employed reaches single phonon sensitivity. We are optimistic that single phonon sensitivity can be achieved in the near future if research effort is made in this direction. In particular, it seems that a precision of tens of phonons is achieved in experiments like the ones presented in \cite{Steinhauer:2014dra} and \cite{Schley:2013pla}. Single phonon sensitivity may be achievable as experimental procedures evolve.
There is a variety of possibilities to measure phonons in BECs, some of them may, in the future, give a precision high enough for our purposes. In particular, one can either try to measure the phase or one can try to measure the density, which are conjugate variables and contain the same information. An interesting approach for measuring the phase is presented in \cite{Katz:2004hig}. It is denoted as ``heterodyne detection" by the authors: After the modes are excited, the trap potential is switched off and the BEC starts to expand and fall freely. During the expansion the energy contained in the phonons is transformed into the kinetic energy of atoms. These free particles interfere with the atoms in the ground state. The interference fringes contain the information about the phonons. Numerical simulations and an approximate analytical derivation for this process are given in \cite{Tozzo:2004pho}. 

Another option for a detection scheme would be time of flight measurements, where phonons are mapped to horizontal atomic momenta and, after a certain time of vertical free fall in the gravitational field of the earth, the momenta can be read from the horizontal position of the atoms. A third option for the measurement of phonons in BECs would be direct light phonon couplings. For example in \cite{Wade:2015squ}, stroboscopic measurements of phonon modes were considered for the creation of squeezing and entanglement of phonon modes. In \cite{Andrews:1996dir,Stamper:1998col} non-destructive phase-contrast imaging was used to observe the bulk perturbations of a BEC. Finally, a fourth option for the measurement of phonons would be the coupling to atomic quantum dots submersed in the BEC \cite{Bruderer:2006prob}. It would be very interesting if experiments could be performed to investigate the sensitivity of different measurement schemes for phonons. A first step towards an experimental realization of our proposal could be experiments using one of the above techniques to simply measure the thermal spectrum of phonons in a BEC with high precision. A second step could be to create phonons in the BEC by Bragg scattering of laser pulses or by periodic modulations of the trapping potential and try to measure them on top of the thermal spectrum. In a last step, the interaction with an oscillating source mass can be implemented.

Additional noise sources that we did not discuss in the main part of this article are Newtonian and seismic noise that give rise to acceleration noise $a_{\mathrm{noise}}^x$. The Newtonian noise also introduces a noise term into the gravity gradient, which we assume to be negligible since the sources of gravity gradient noise will be far away in comparison to the extension of the BEC, which means that the gravity gradient noise will be highly suppressed in comparison to the gravitational acceleration noise \footnote{This is a clear advantage of measurements of the gravity gradient of small objects besides the property that the gravity gradient of a sphere close to its surface is independent of its radius and only depends on its mass.}. 
A generic example of the square root of the displacement spectral density, $S^{1/2}_x$, in a modern laboratory environment close to traffic is shown in Fig. 3.3 of Tobias Westphal's PhD thesis \cite{Westphal:2016coa} for the case of the physics department of the university of Hannover. The square root of the displacement spectral density $S^{1/2}_x$ for $1\rm{Hz}$ is of the order $10^{-7}\rm{m\,Hz^{-1/2}}$  \footnote{Actually, there is a dip slightly below $1\,\rm{Hz}$, which the experimental parameters may be tuned into to lower the seismic noise background.}. We can assume that the laboratory structure is not driven resonantly in this frequency range and that damping can be neglected. Then, the susceptibility can be approximated as $1/\omega^2$, and we find an acceleration spectral density of about $S^{1/2}_{a^x}=\omega^2 S^{1/2}_x \sim 10^{-6}\,\rm{ms^{-2}\,Hz^{-1/2}}$ at $\omega=2\pi\times 1\,\rm{Hz}$. After $t_\rm{exp}=10\,\rm{s}$ and $\#_\rm{rep}\sim 10^4$, this leads to $a_\rm{min}(\omega_{1}) = S^{1/2}_{a^x}/\sqrt{t_\rm{int}\#_\rm{rep}} \sim 10^{-8}\,\rm{ms^{-2}}$. Hence, the Newtonian and seismic noise background has to be lower by only about one order of magnitude to get below the order of the gravitational acceleration due to a $200\,\rm{g}$ source mass. For the case of $M=200\,\rm{mg}$, the Newtonian and seismic noise have to suppressed by two orders. Both situations should be achievable by choosing a quieter environment, e.g. a site underground far from human induced noise, and a vibration isolation chain \cite{Schmoele:2017diss}. As an example, advanced LIGO is engineered to achieve noise levels at the mirrors of about $5\times 10^{-19}\rm{m\,Hz^{-1/2}}$ \cite{Martynov:2016fzi} for frequencies above $10\,\rm{Hz}$, which would be more than sufficient to make the acceleration noise negligible in comparison to the gravitational acceleration induced by the source mass. In particular, vibration isolation chains have to be included in any case since the source and the detector must not be coupled significantly through the devices holding them at their respective positions.

\begin{acknowledgments}
We thank Florent Michel, David Edward Bruschi, J\"org Schmiedmayer, Michael Gring, Jonas Schm\"ole, Tobias Westphal, Philipp Haslinger, Philipp Thomas, Renaud Parentani, Luis Cort\'es Barbado, Jan Kohlrus and Ana Luc\'ia Baez for interesting remarks and discussions. DR thanks the Humboldt Foundation for funding his research in Vienna with their Feodor-Lynen Felloship  and Kiri Mochrie for writing assistance. R.H. and I.F. would like to acknowledge the support of the grant ``Leaps in
cosmology: gravitational wave detection with quantum
systems'' (No. 58745) from the John Templeton Foundation. The publication of this article is funded by the Open Access Publishing Fund of the University of Vienna. 
\end{acknowledgments}

\appendix

\section{The perturbations due the stationary part of the gravitational potential}
\label{sec:stationary}

Let us find solutions of Eq. (\ref{eq:bulkpert1}) and Eq. (\ref{eq:bulkpert2}) when only the time-independent part of the gravitational potential is considered. We call these solutions $\alpha^\rm{c}$ and $\phi^\rm{c}$ and we only consider terms of first order in $\alpha^\rm{c}$, $\phi^\rm{c}$ and the gravitational potential. Then, we have to set $\dot \alpha^c=0$ in Eq. (\ref{eq:bulkpert2}), and we find that $\phi^{\rm{c}\prime\prime}=0$ inside the trap and that $\phi^\rm{c}$ has to vanish at the potential walls. Therefore, it follows from $\phi^{\rm{c}\prime\prime}=0$ that $\phi^{\rm{c}\prime}$ has to vanish everywhere and $\phi^\rm{c}=\phi^\rm{c}(t)$. From Eq. (\ref{eq:bulkpert1}), we find that $\alpha^{\rm{c}\prime}$ has to vanish at the boundaries. Only one equation remains for the stationary density perturbation inside the trap
\begin{equation}\label{eq:bulkpert1.1}
	-\dot\phi^\rm{c} = - \frac{\hbar}{2m}\alpha^{\rm{c}\,\prime\prime} + \frac{\hbar}{m}\lambda \rho_0 \alpha^\rm{c}  + \frac{m}{\hbar} \Phi^\rm{c}\,,
\end{equation}
Since the right hand side of Eq. (\ref{eq:bulkpert1.1}) is time independent, $\dot\phi^\rm{c}$ must be a constant. We define the perturbation of the chemical potential $\delta\mu^\rm{c} := -\hbar \dot\phi^\rm{c}$. Using again the Thomas-Fermi approximation, we neglect the kinetic energy term in Eq. (\ref{eq:bulkpert1.1}) everywhere up to a small region at the boundary of length $\zeta$. Then, we obtain
\begin{equation}\label{eq:bulkpert1.2}
	\frac{\delta\mu^\rm{c}}{\hbar} = \frac{\hbar}{m}\lambda \rho_0 \alpha^\rm{c} + \frac{m}{\hbar} \Phi^\rm{c}\,.
\end{equation}
which leads to 
\begin{equation}
	\alpha^\rm{c} \approx \frac{1}{2 m c_0^2}\left(-m \Phi^\rm{c} + \delta\mu^\rm{c}  \right)\,.
\end{equation}
up to a region of length $\zeta$ at the boundaries of the trap potential. Additionally, $\alpha^\rm{c}$ has to fulfill the condition $\int_{-L/2}^{L/2}dx\,\alpha^\rm{c}=0$ since the total number of atoms $N\approx \mathcal{A}\int_{-L/2}^{L/2}dx\, \rho_0(1+2\alpha^\rm{c})$ is conserved. We obtain the expression for the perturbation of the chemical potential
\begin{eqnarray}\label{eq:deltamu}
	\delta\mu^\rm{c} &=& \frac{m}{L} \int_{-L/2}^{L/2}dx\, \Phi^\rm{c}(x) = \frac{m}{L} \int_{-L/2}^{L/2}dx\,\left(\Phi^c_0 + a^c x + \mathfrak{G}^{c} \frac{x^2}{2}\right)=m\Phi^c_0 + \frac{m}{24} \mathfrak{G}^{c} L^2\,,
\end{eqnarray}
and finally
\begin{equation}
	\alpha^\rm{c} \approx \frac{1}{2 c_0^2}\left(- a^c x + \frac{\mathfrak{G}^{c}}{2} \left(\frac{1}{12} L^2 - x^2 \right)  \right)\,,
\end{equation}
up to a small region close to the boundary in which $\alpha''$ cannot be neglected and enables $\alpha'=0$ to vanish at the boundary

Now, we have to check if the equations (\ref{eq:bulkpert1}) and (\ref{eq:bulkpert2}) are still approximately correct if we replace $\sqrt{\rho_0}$ by $\sqrt{\tilde \rho_0}=\sqrt{\rho_0}(1+\alpha^c)$.  We assume that the number of atoms in the BEC is of the order $4\times 10^{6}$, the length of the BEC is $L=300\rm{\mu m}$ and its density is $\rho_0\approx 10^{13}\,\rm{cm}^{-3}$. For a ${}^{87}\rm{Rb}$-BEC we have a self-interaction constant $\lambda\approx 10^{-7}\rm{m}$. If we assume an average distance $R_\rm{min}$ of $1\,\rm{mm}$ between the center of the trapping potential and the center of a $200\,\rm{g}$ tungsten/gold mass, we find that $|\alpha^c|$ has its maximum of the order of $10^{-6}$ at $x=\pm L/2$. The perturbation of the time derivative of the phase $\dot\phi^\rm{c}=-\delta\mu/\hbar$ can be compared with $\dot \theta_0=-\mu/\hbar$. We find that $\dot\phi^\rm{c}/\dot\theta_0\sim 10^{-8}$. Therefore, all terms proportional to $\alpha^c$ and $\phi^c$ that can appear in Eq. (\ref{eq:bulkpert1}) and Eq. (\ref{eq:bulkpert2}) are negligible, and we are justified to treat the effect of the sinusoidally time-dependent terms in $\Phi$ independently of $\alpha^c$ and $\phi^c$.

\section{Phenomenological treatment of dissipation}
\label{sec:dissipation}

The basic idea of the phenomenological treatment of dissipation presented in \cite{Pitaevskii:1959phen} and \cite{Choi:1998ia} is that there exists an equilibrium state $\psi_0$ that fulfills the undamped, time-independent Gross-Pitaevskii equation $\mu \psi_0 = \mathcal{H}[|\psi_0|^2]\, \psi_0= (-\hbar^2/2m\,\, \partial_1^2  + \lambda \hbar^2 |\psi_0|^2/2m + V)\,\psi_0$ for a given chemical potential $\mu$. Therefore, the time evolution operator for the damped system out of equilibrium has to vanish identically on this state. This operator is derived by removing the chemical potential from the differential operator $\mathcal{H}[|\psi_0|^2]$ and multiplying the resulting differential operator with the factor $(1+i\Lambda)$, where $\Lambda$ is the damping constant. Then, the resulting non-unitary differential operator vanishes on the equilibrium state as wanted.

From the Gross-Pitaevskii equation, we obtain the time independent Gross-Pitaevskii equation when $i\hbar\partial_t$ is replaced by $\mu$. The equilibrium state is the initial state $\psi_0=\sqrt{\rho_0}e^{i\theta_0}$, which we introduced in Sec. \ref{sec:descriptionfluid} for $V=0$, and we obtain the chemical potential $\mu=\hbar^2\lambda \rho_0/2m$. Through the substitutions $\mathcal{\tilde H} = \mathcal{H}-\mu$ and $\tilde \psi = \psi e^{i\mu t/\hbar}$, where $\mu t/\hbar=-\theta_0$, and the multiplication of $\mathcal{\tilde{H}}$ with $(1+i\Lambda)$ we arrive at the damped equation. Now, we introduce the external potential $V=\Phi$ that drives the system and we find the damped, driven equation $i\hbar\partial_t\tilde\psi=(1+i\Lambda)\mathcal{\tilde{H}}[|\tilde\psi|^2]\,\tilde\psi+V\tilde\psi$.
For very small $\Lambda$, multiplying $\mathcal{\tilde H}$ with $(1+i\Lambda)$ is equivalent to multiplying the time derivative $i\hbar\partial_t$ and $V$ with $(1-i\Lambda)$ which gives 
\begin{equation}\label{eq:GPnonlineardamped}
i\hbar(1-i\Lambda)\partial_t\tilde\psi=\mathcal{\tilde{H}}[|\tilde\psi|^2]\,\tilde\psi+(1-i\Lambda)V\tilde\psi\,.
\end{equation}
Following the steps in Sec. \ref{sec:descriptionfluid}, we obtain Eq. (\ref{eq:harmoscdamp2}) in first order in $\Lambda$.

\section{The interaction Hamiltonian}
\label{sec:intham}

We start from the Hamiltonian for a gas of interacting Bosons in an external potential $\mathcal{V}_\rm{ext}$ \cite{Pitaevskii:2003bose}
\begin{eqnarray}
	\hat H &=& \int d^3x\, \hat\Psi^\dagger\left(-\frac{\hbar^2}{2m} \nabla^2  + \mathcal{V}_\rm{ext}\right)\hat\Psi + \frac{g}{2} \int d^3x  \,\hat\Psi^\dagger \hat\Psi^\dagger \hat\Psi \hat\Psi\,.
\end{eqnarray}
where $g=\hbar^2\lambda/(2m)$ and it is assumed that $\hat\Psi$ and $\hat\Psi^\dagger$ vanish at the boundaries of integration. We split the potential in a time-independent part and a time-dependent perturbation as $\mathcal{V}_\rm{ext}=\mathcal{V}_{0\rm{ext}} + \delta\mathcal{V}_\rm{ext}$ such that
\begin{eqnarray}
	\hat H &=& \int d^3x\, \hat\Psi^\dagger\left(-\frac{\hbar^2}{2m} \nabla^2  + \mathcal{V}_{0\rm{ext}}\right)\hat\Psi + \int dx\, \hat\Psi^\dagger \delta\mathcal{V}_\rm{ext}\hat\Psi  + \frac{g}{2} \int d^3x\, \hat\Psi^\dagger \hat\Psi^\dagger \hat\Psi \hat\Psi\,.
\end{eqnarray} 
We are working in the Heisenberg picture, where $\Psi$ is time dependent and the states are constant. We assume the equal time commutation relation of the field operator and its complex conjugate $[\hat\Psi(z,t),\hat\Psi^\dagger(z',t)]=\delta(z-z')$ and $[\hat\Psi(z,t),\hat\Psi(z',t)]=0$. The Hamiltonian $\hat H$ governs the evolution of the field operator via the Heisenberg equation of motion $-i\hbar\partial_t \hat\Psi = [\hat H,\hat\Psi]$. 

We define the normalized ground state wave function of the BEC $\bar\psi_0(z)$ as the solution of the stationary Gross-Pitaevskii (PT) equation
\begin{equation}\label{eq:statgp}
	\left(-\frac{\hbar^2}{2m} \nabla^2  + \mathcal{V}_{0\rm{ext}}  + gN_a|\bar\psi_0|^2 \right)\bar\psi_0 = \mu \bar\psi_0\,,
\end{equation}
where $\mu$ is the chemical potential and $N_a$ is the number of atoms in the BEC ground state. Furthermore, we define a variation of the chemical potential $\delta\mu(t)$ as the normalized moment $\delta\mu(t):= \int d^3x\, \bar\psi_0^*\delta\mathcal{V}_\rm{ext}\bar\psi_0/\int d^3x\, \bar\psi_0^*\bar\psi_0$. Then, $\bar\psi'_0:=\bar\psi_0e^{-i\int_0^t dt'\,\delta\mu(t')/\hbar}$ solves the time dependent GP equation
\begin{equation}
	\left(-\frac{\hbar^2}{2m} \nabla^2  + \mathcal{V}_{0\rm{ext}} + \delta\mu(t) + gN_a|\bar\psi_0|^2 \right)\bar\psi'_0 = i\hbar\partial_t \bar\psi'_0\,.
\end{equation}
We consider the expansion 
\begin{eqnarray}
	\hat\Psi(z,t)=\hat\Psi'(z,t)e^{-i\mu t/\hbar - i\int_0^t dt'\,\delta\mu(t')/\hbar}=(\hat\Psi_0(z) + \hat\vartheta(z,t))e^{-i\mu t/\hbar - i\int_0^t dt'\,\delta\mu(t')/\hbar}
\end{eqnarray} 
in the Heisenberg picture, where $\hat\Psi_0 = \hat a_0 \bar\psi_0$. Since the equal time commutation relations are the same for $\hat\Psi'$ as for $\hat\Psi$, we find that the time evolution of $\hat\Psi'$ is governed by the grand canonical Hamiltonian $\hat H':= \hat H - (\mu+\delta\mu) \hat N$ via the Heisenberg equation of motion $-i\hbar\partial_t \hat\Psi' =  [\hat H',\hat\Psi']$, where $\hat N(t) = \int d^3x\, \hat\Psi^{\prime\dagger}(z,t) \hat\Psi^\prime(z,t)$.

We assume that the state of the lowest energy mode can be considered as a coherent state $|\alpha_0\rangle$ with $\alpha_0=\sqrt{N_a}=:1/\xi\gg 1$. Then $\hat a_0 |\alpha_0\rangle = \alpha_0 |\alpha_0\rangle$ and $\hat a_0^\dagger |\alpha_0\rangle \approx \alpha_0 |\alpha_0\rangle$, and we can replace the operator $\hat\Psi_0(z)$ with the function $\Psi_0(z)=\alpha_0 \bar\psi_0(z)$. Using $\hat\Psi(z,t)=\xi^{-1}(\bar\psi_0(z) + \xi\hat\vartheta(z,t))e^{-i\mu t/\hbar - i\int_0^t dt'\,\delta\mu(t')/\hbar}$
in $\hat H$, we can write
\begin{eqnarray}\label{eq:expansionfull}
	\nonumber  \hat H' &=& \xi^{-2}\int d^3x\, \bar\psi_0^{*}\left(-\frac{\hbar^2}{2m} \nabla^2  + \mathcal{V}_{0\rm{ext}} + \frac{gN_a}{2}|\bar\psi_0|^2\right)\bar\psi_0  \\
	\nonumber && + \xi^{-1}\int d^3x\, \left(\hat\vartheta^{\dagger}\left(-\frac{\hbar^2}{2m} \nabla^2  + \mathcal{V}_{0\rm{ext}} + gN_a|\bar\psi_0|^2 \right)\bar\psi_0 + h.c.\right) \\
	\nonumber && + \int d^3x\, \left(\hat\vartheta^{\dagger}\left(-\frac{\hbar^2}{2m} \nabla^2  + \mathcal{V}_{0\rm{ext}} + 2gN_a|\bar\psi_0|^2 \right)\hat\vartheta + \frac{gN_a}{2}\left(\hat\vartheta^{\dagger 2}\bar\psi_0^{ 2} + \bar\psi_0^{*2}\hat\vartheta^{2}\right)\right)
	\\
	&& + \xi^{1}\int d^3x\, gN_a\left(\hat\vartheta^{\dagger 2}\hat\vartheta \bar\psi_0 + \bar\psi_0^{*}\hat\vartheta^{\dagger}\hat\vartheta^{2}\right) + \xi^{2}\int d^3x\, \frac{gN_a}{2}\hat\vartheta^{\dagger 2}\hat\vartheta^{2}\\
	\nonumber && + \xi^{-2}\int d^3x\, \bar\psi_0^{*} \delta\mathcal{V}_\rm{ext}\bar\psi_0  + \xi^{-1}\int d^3x\, \left(\bar\psi_0^{*} \delta\mathcal{V}_\rm{ext}\hat\vartheta + \hat\vartheta^{\dagger} \delta\mathcal{V}_\rm{ext}\bar\psi_0\right) +  \int d^3x\, \hat\vartheta^{\dagger} \delta\mathcal{V}_\rm{ext}\hat\vartheta \\
	\nonumber && - \xi^{-2}\mu \int d^3x\, \bar\psi_0^{*} \bar\psi_0 - \xi^{-2}\delta\mu \int d^3x\, \bar\psi_0^{*} \bar\psi_0 - \xi^{-1}\mu \int d^3x\, \left(\bar\psi_0^{*} \vartheta + \vartheta^{\dagger}\bar\psi_0\right)  \\
	\nonumber &&  - \xi^{-1}\delta\mu \int d^3x\, \left(\bar\psi_0^{*} \vartheta + \vartheta^{\dagger}\bar\psi_0\right) - \mu\int d^3x\, \hat\vartheta^{\dagger} \hat\vartheta- \delta\mu\int d^3x\, \hat\vartheta^{\dagger} \hat\vartheta
\end{eqnarray}
In the last three lines, we see the contribution of the time dependent potential perturbation and $ - (\mu+\delta\mu) \hat N$. We find that the second term in the second last line and the first term in the third last line cancel. 
With Eq. (\ref{eq:statgp}), the first line in Eq. (\ref{eq:expansionfull}) gives the classical energy of the condensate, and with the first term in the second last line of Eq. (\ref{eq:expansionfull}), we find
\begin{equation}
	\hat H^{(0)} =  E^{(0)} = -\frac{gN_a^2}{2}\int d^3x\, |\bar\psi_0|^4\,.
\end{equation} 
Again with the stationary GP equation (\ref{eq:statgp}), the second line of Eq. (\ref{eq:expansionfull}) becomes
\begin{equation}
	\hat H^{(1)} =  \sqrt{N_a}\mu\int d^3x\, \left(\hat\vartheta^{\dagger}\bar\psi_0 + h.c.\right)\,,
\end{equation} 
which cancels with the last term in the second last line of Eq. (\ref{eq:expansionfull}). The third line of Eq. (\ref{eq:expansionfull}) gives rise to the Bogoliubov Hamiltonian. We combine the third line and the second term in the last line of Eq. (\ref{eq:expansionfull}) as
\begin{eqnarray}
	\hat H^{(2)} = \mathcal{H}^{(2)}[\hat\vartheta] := \quad : && \int d^3x\, \left(\hat\vartheta^{\dagger}\left(-\frac{\hbar^2}{2m} \nabla^2  + \mathcal{V}_{0\rm{ext}} - \mu + 2gN_a|\bar\psi_0|^2 \right)\hat\vartheta \right. \\
	&& \left. + \frac{gN_a}{2}\left(\hat\vartheta^{\dagger 2}\bar\psi_0^{2} + \bar\psi_0^{*2}\hat\vartheta^{2}\right)\right)\,: \,,
\end{eqnarray} 
where $:\quad:$ denotes the normal ordering, which leads to the omission of the constant vacuum energy.
Furthermore, we combine the remaining terms of Eq. (\ref{eq:expansionfull}) to the interaction Hamiltonian
\begin{eqnarray}\label{eq:Hprimeintappend}
	  \hat H_\rm{int} = \mathcal{H}_\rm{int}[\hat\vartheta] &:=&  \xi^{1}gN_a\int d^3x\, \left(\hat\vartheta^{\dagger 2}\hat\vartheta \bar\psi_0 + \bar\psi_0^{*}\hat\vartheta^{\dagger}\hat\vartheta^{2}\right) + \xi^{2}\frac{gN_a}{2}\int d^3x\, \hat\vartheta^{\dagger 2}\hat\vartheta^{ 2}\\
	\nonumber &&  + \xi^{-1}\int d^3x\, (\delta\mathcal{V}_\rm{ext}-\delta\mu)\left(\bar\psi_0^{*} \hat\vartheta + \hat\vartheta^{\dagger} \bar\psi_0\right) +  \int d^3x\, \hat\vartheta^{\dagger}(\delta\mathcal{V}_\rm{ext}-\delta\mu)\hat\vartheta\,.
\end{eqnarray}
The split $\sqrt{N_a} \bar\psi_0(x) + \hat\vartheta(x)$ corresponds to the initial split at $t_0$ before the interaction with the external potential is switched on.
Therefore, we can assume that the field operator $\hat\vartheta(x)$ only contains ladder operators of the modes $n>0$ and we can write 
\begin{equation}
	\hat\vartheta(x) = \sum_n\left(u_n(x)\hat b_{n} + v_n^*(x)\hat b_{n}^\dagger\right)\,,
\end{equation}
where $[b_{n},b_{m}^\dagger]=\delta_{nm}$ and the mode functions fulfill the stationary Bogoliubov-deGennes (BDG) equations
\begin{eqnarray}\label{eq:BDG}
	\hbar \omega_n u_n(x) &=& \left(-\frac{\hbar^2}{2m} \nabla^2  + \mathcal{V}_{0\rm{ext}} - \mu + 2gN_a|\bar\psi_0|^2\right)u_n(x) + gN_a\bar\psi_0^2 v_n(x)\\
	-\hbar \omega_n v_n(x) &=& \left(-\frac{\hbar^2}{2m} \nabla^2  + \mathcal{V}_{0\rm{ext}} - \mu + 2gN_a|\bar\psi_0|^2\right)v_n(x) + gN_a\bar\psi_0^{*2} u_n(x)\,.
\end{eqnarray}
Furthermore, we assume that the solutions $u_n$ and $v_n$ are ortho-normalized as
\begin{equation}\label{eq:normauvappend}
	\int d^3x\, \left(u_n^*(x) u_m(x) - v_n^*(x) v_m(x)\right) = \delta_{nm}\,.
\end{equation}
We obtain for the normal ordered free Hamiltonian
\begin{eqnarray}
	\nonumber \hat H^{(2)} &=&  \sum_n  \,\hbar \omega_n \hat b_{n}^\dagger \hat b_{n}\,.
\end{eqnarray}

\section{Damping in 3-dimensional box traps}
\label{sec:dampbox}

In this appendix, we show that damping of phonons in a BEC in a uniform box trap does not differ significantly from phonon damping in a uniform BEC with periodic boundary conditions in the parameter range that we consider in this article. From Eqs. (37), (39) and (40) of \cite{Giorgini:1997sca}, an explicit expression for the damping rate in a BEC can be given as
\begin{eqnarray}\label{eq:damprateexpl}
	\gamma_n &=& 4\pi g^2\sum_{ij} (f^0_i - f^0_j) |A^n_{ij}|^2 \delta(\hbar (\omega_0 + \omega_i - \omega_j)) \\
	&& + 2\pi g^2 \sum_{ij} (1 + f^0_i + f^0_j) \left(|B^n_{ij}|^2 \delta(\hbar (\omega_0 - \omega_i - \omega_j)) - |\tilde{B}^n_{ij}|^2 \delta(\hbar (\omega_0 + \omega_i + \omega_j)\right)\,, 
\end{eqnarray}
where 
\begin{eqnarray}\label{eq:dampmom}
	A^n_{ij}&=&\int d^3x \,\phi_0 \left[ u_n(u_i u_j^* + v_i v_j^* + v_i u_j^*) + v_n(u_i u_j^* + v_i v_j^* + u_i v_j^*)\right]\\
	B^n_{ij}&=&\int d^3x \,\phi_0  \left[ u_n(u_i^* v_j^* + v_i^* u_j^* + u_i^* u_j^*) + v_n(u_i^* v_j^* + v_i^* u_j^* + v_i^* v_j^*)\right]\\
	\tilde{B}^n_{ij}&=&\int d^3x \,\phi_0 \left[ u_n(u_i v_j + v_i u_j + v_i v_j) + v_n(u_i v_j + v_i u_j + u_i v_j)\right]\,.
\end{eqnarray}
where $\phi_0=\sqrt{\rho_0}$ and the indeces are multi-component, i.e. $i=(i_x,i_y,i_z)$ and so on. We are considering interaction times that are only a factor 5 larger than the inverse frequency of the modes under consideration in this article. Therefore, the delta function in Eq. (\ref{eq:damprateexpl}) have to be replaced by a sinc-function as \cite{Gotlibovych:2014deg}
\begin{equation}\label{eq:sinc}
	\delta(\hbar(\omega_n \pm \omega_i \pm \omega_j))=\frac{t_\rm{exp}}{2\pi\hbar}\,\rm{sinc}^2\left(\frac{t_\rm{exp}}{2}(\omega_n \pm \omega_i \pm \omega_j)\right)\,.
\end{equation}
The width of the sinc function $\rm{sinc}(t_\rm{exp}(\omega_0 + \omega_i + \omega_j)/2)$ is still small enough to justify that we can neglect the third term in Eq. (\ref{eq:damprateexpl}).
Let us assume that the BEC trap has a square cross section of edge length $L_\rm{tr}$. In analogy to the expressions for the phonon mode functions for a BEC in a box in the $x$-direction given in Eq. (\ref{eq:modefunctionsuv}), we can solve the three dimensional Bogoliubov-DeGennes equations in a box trap using the mode functions
\begin{eqnarray}
	u_{n} &=& \alpha_n\cos{k_{n_x} (x + L/2)}\cos{k_{n_y} (y + L_\rm{tr}/2)}\cos{k_{n_z} (z + L_\rm{tr}/2)}\quad\rm{and}\\
	v_n &=& \beta_n\cos{k_{n_x} (x + L/2)}\cos{k_{n_y} (y + L_\rm{tr}/2)}\cos{k_{n_z} (z + L_\rm{tr}/2)}
\end{eqnarray}
where $\alpha_n = (4c_0(\sqrt{2}\zeta \omega_n)^{-1} + 1)^{1/2}(L L_\rm{tr}^2)^{-1/2}$, $\beta_n = -(4c_0(\sqrt{2}\zeta \omega_n)^{-1} - 1)^{1/2}(L L_\rm{tr}^2)^{-1/2}$, $k_{n_x}=n_x\pi/L_\rm{tr}$, $k_{n_y}=n_y\pi/L_\rm{tr}$, $k_{n_z}=n_z\pi/L$ and $\omega_n=c_0(k_{n_x}^2+k_{n_y}^2+k_{n_z}^2)^{1/2}$. We are interested in the particular case where $n_x=0=n_y$. Then, we find for the remaining moments in Eq. (\ref{eq:dampmom})
\begin{eqnarray}\label{eq:dampmombox}
	A^n_{ij}=\bar{A}^n_{ij}  M^n_{ij} \quad\rm{and}\quad B^n_{ij} = \bar{B}^n_{ij}  M^n_{ij} \,,
\end{eqnarray}
where 
\begin{equation}
	M^n_{ij} = \left( \delta_{i_x,j_x} + \delta_{i_x,0}\delta_{j_x,0}\right)\left( \delta_{i_y,j_y} + \delta_{i_y,0}\delta_{j_y,0}\right)\left( \delta_{i_z-j_z,n_z} + \delta_{i_z+j_z,n_z} + \delta_{j_z-i_z,n_z} \right)
\end{equation}
and $\bar{A}^n_{ij}$ and $\bar{B}^n_{ij}$ can be approximated as
\begin{eqnarray}\label{eq:dampmomfac}
	\bar{A}^n_{ij}&\approx& \frac{\rho_0^2g^2(\omega_n + \omega_i - \omega_j) + \frac{3}{4} \hbar^2\pi^2  \omega_i\omega_j\omega_n}{2\sqrt{\pi} \sqrt{L L_\rm{tr}^2} \rho_0 g^{3/2} (\hbar\omega_i\omega_j\omega_n)^{1/2}}  \\
	\bar{B}^n_{ij}&\approx& \frac{-\rho_0^2g^2(\omega_n - \omega_i - \omega_j) + \frac{3}{4} \hbar^2\pi^2  \omega_i\omega_j\omega_n}{2\sqrt{\pi} \sqrt{L L_\rm{tr}^2} \rho_0 g^{3/2} (\hbar\omega_i\omega_j\omega_n)^{1/2}}\,.
\end{eqnarray}
The sinc-functions in Eq. (\ref{eq:damprateexpl}) only deliver significant contributions in combination with $M^n_{ij}$ when the momentum relation agrees with the energy relation in the argument of the sinc-function. In other words
\begin{eqnarray}
	A^n_{ij} &=&\bar{A}^n_{ij} \left( \delta_{i_x,j_x} + \delta_{i_x,0}\delta_{j_x,0}\right)\left( \delta_{i_y,j_y} + \delta_{i_y,0}\delta_{j_y,0}\right) \delta_{j_z-i_z,n_z}\quad\rm{and}\\
	B^n_{ij} &=& \bar{B}^n_{ij} \left( \delta_{i_x,j_x} + \delta_{i_x,0}\delta_{j_x,0}\right)\left( \delta_{i_y,j_y} + \delta_{i_y,0}\delta_{j_y,0}\right) \delta_{i_z+j_z,n_z} \,.
\end{eqnarray}
For $n_z$ of the order of 1, the relation $i_z+j_z=n_z$ in combination with $i_x=j_x$ or $i_y=j_y$ will lead to a value for $\omega_n - \omega_i - \omega_j$ much larger than the width of the sinc-function. Therefore, we can write 
\begin{eqnarray}
	B^n_{ij} &=& 4\bar{B}^n_{ij}\delta_{i_x,0}\delta_{j_x,0}\delta_{i_y,0}\delta_{j_y,0}\delta_{i_z+j_z,n_z} \,.
\end{eqnarray}
Let us compare these results with the corresponding expressions for the uniform BEC with periodic boundary conditions. We use the normalized mode functions
\begin{eqnarray}
	u_{n} &=& \frac{\alpha_n}{\sqrt{8}}\exp{ik_{n_x}x}\exp{ik_{n_y} y}\exp{ik_{n_z}z}\quad\rm{and}\\
	v_n &=& \frac{\beta_n}{\sqrt{8}}\exp{ik_{n_x}x}\exp{ik_{n_y}y}\exp{ik_{n_z}z}\,,
\end{eqnarray}
where $k_{n_x}=2n_x\pi/L_\rm{tr}$, $k_{n_y}=2n_y\pi/L_\rm{tr}$, $k_{n_z}=2n_z\pi/L$ and $\alpha_n$ and $\beta_n$ are defined through $k_{n_x}$, $k_{n_y}$ and $k_{n_z}$ as above. For $n_x=0=n_y$, we obtain
\begin{eqnarray}\label{eq:dampmomuniform}
	|A^n_{ij}|&=&\frac{1}{\sqrt{2}}\bar{A}^n_{ij}\delta_{j_x,i_x}\delta_{j_y,i_y}\delta_{j_z-i_z,n_z} \,, \\
	 |B^n_{ij}| &=&\frac{1}{\sqrt{2}}\bar{B}^n_{ij}  \delta_{j_x,0}\delta_{i_x,0}\delta_{j_y,0}\delta_{i_y,0} \quad \rm{and} \quad |\tilde{B}^n_{ij}| = 0\,.
\end{eqnarray}
We find that the Landau damping rate for phonons in a BEC in a box potential is of the same order as the Landau damping rate in a uniform BEC with periodic boundary conditions. For the Beliaev damping, we find an increase of about one order. Since Beliaev damping is strongly suppressed for our parameter range, we conclude that we can use the expressions for the damping constants derived for periodic boundary conditions to describe the BEC in a box potential.

\bibliography{milligrammass}

\end{document}